\documentclass[a4paper, 10pt, leqno]{amsart}

\usepackage{amsmath,amsfonts}
\usepackage{amssymb}
\usepackage{amsmath}
\usepackage{enumitem}
\usepackage[fleqn]{mathtools}
\usepackage{graphicx}
\usepackage{epstopdf}
\usepackage{rotating}
\usepackage{caption}
\usepackage{fancyhdr}
\usepackage{float}
\usepackage{hyperref}
\usepackage{subcaption}
\usepackage{booktabs}
\usepackage[foot]{amsaddr}
\usepackage{dsfont}

\usepackage[utf8]{inputenc}
\usepackage[T1]{fontenc}
\usepackage[autolanguage]{numprint}

\usepackage{xcolor}
\usepackage{mdwlist}
\usepackage{amsaddr}

\hypersetup{
  colorlinks   = true, 
  urlcolor     = blue, 
  linkcolor    = blue, 
  citecolor   = red 
}

\begin{document}
\pagestyle{plain}
\bibliographystyle{plain}
\newtheorem{theo}{Theorem}[section]
\newtheorem{lemme}[theo]{Lemma}
\newtheorem{cor}[theo]{Corollary}
\newtheorem{defi}[theo]{Definition}
\newtheorem{prop}[theo]{Proposition}
\newtheorem{remarque}[theo]{Remark}
\newtheorem{claim}[theo]{Claim}
\newcommand{\beq}{\begin{eqnarray}}
\newcommand{\enq}{\end{eqnarray}}
\newcommand{\be}{\begin{eqnarray*}}
\newcommand{\en}{\end{eqnarray*}}
\newcommand{\ben}{\begin{eqnarray*}}
\newcommand{\enn}{\end{eqnarray*}}
\newcommand{\Td}{\mathbb T^d}
\newcommand{\Rd}{\mathbb R^n}
\newcommand{\R}{\mathbb R}
\newcommand{\N}{\mathbb N}
\newcommand{\Sn}{\mathbb S}
\newcommand{\Zd}{\mathbb Z^d}
\newcommand{\Linf}{L^{\infty}}
\newcommand{\dt}{\partial_t}
\newcommand{\Dt}{\frac{d}{dt}}
\newcommand{\Dtt}{\frac{d^2}{dt^2}}
\newcommand{\demi}{\frac{1}{2}}
\newcommand{\vf}{\varphi}
\newcommand{\epu}{_{\varepsilon}}
\newcommand{\ep}{^{\varepsilon}}
\newcommand{\bfi}{{\mathbf \Phi}}
\newcommand{\bpsi}{{\mathbf \Psi}}
\newcommand{\bx}{{\mathbf x}}
\newcommand{\bC}{{\mathbf C}}
\newcommand{\dis}{\displaystyle}
\newcommand{\ds}{\partial_s}
\newcommand{\dss}{\partial_{ss}}
\newcommand{\dx}{\partial_x}
\newcommand{\dxx}{\partial_{xx}}
\newcommand{\dy}{\partial_y}
\newcommand{\dyy}{\partial_{yy}}
\newcommand {\g}{\`}
\newcommand{\E}{\mathbb E}
\newcommand{\bQ}{\mathbb Q}
\newcommand{\1}{\mathbb I}
\newcommand{\F}{\cal F}
\let\cal=\mathcal
\newcommand{\lb}{\langle}
\newcommand{\rb}{\rangle}
\newcommand{\bv}{\bar{v}}
\newcommand{\uv}{\underline{v}}
\newcommand{\cS}{{\cal S}}
\newcommand{\cSf}{\cS_\Phi}
\newcommand{\bw}{\bar{w}}
\newcommand{\uw}{\underline{w}}
\newcommand{\bsig}{\bar{\sigma}}
\newcommand{\usig}{\underline{\sigma}}
\newcommand{\vr}{\mathrm{r}}
\newcommand{\mw}{\mathbf{w}}
\newcommand{\bV}{{\mathbf V}}
\def\red#1{\textcolor[rgb]{1,0,0}{ #1}}
\def\bb#1{\mathbf{#1}}
\def \d#1{\partial_{#1}}

\newcommand{\bi}{\begin{itemize}}
\newcommand{\ei}{\end{itemize}}
\newcommand{\e}{\mathbb{E}}
\newcommand{\et}{\tilde{\mathbb{E}}}
\newcommand{\p}{\mathbb{P}}
\newcommand{\f}{\mathcal{F}}
\newcommand{\ft}{\tilde{\mathcal{F}}}
\newcommand{\fk}{\mathcal{F}_{t_k}}
\renewcommand{\l}{\mathbb{L}}
\newcommand{\h}{\mathbb{H}}
\newcommand{\prog}{\mathcal{P}}
\newcommand{\Pb}{\bold{P}}
\newcommand{\Filt}{\mathcal{F}}

\newenvironment{demo}{\par\bigskip\noindent {\sc Proof: }\normalsize}{\hfill $\Box$ \par \medskip}
\def \dd{{\rm d}}

\def\greg#1{{\color{red}#1}}
\def\red#1{{\color{red}#1}}
\def\blue#1{{\color{blue}#1}}

\newtheorem{theorem}{Theorem}[section]
\newtheorem{lemma}[theorem]{Lemma}
\newtheorem{corollary}[theorem]{Corollary}
\newtheorem{proposition}[theorem]{Proposition}
\theoremstyle{definition}\newtheorem{definition}[theorem]{Definition}
\theoremstyle{definition}\newtheorem{problem}{Problem}
\theoremstyle{definition}\newtheorem{formulation}{Formulation}
\theoremstyle{definition}\newtheorem{example}[theorem]{Example}
\theoremstyle{definition}\newtheorem{assumption}[theorem]{Assumption}
\theoremstyle{definition}\newtheorem{remark}[theorem]{Remark}

\newcommand{\eq}[1]{\begin{align}#1\end{align}}
\newcommand{\eqn}[1]{\begin{align*}#1\end{align*}}

\newcommand{\bE}{\mathbb E}
\newcommand{\bB}{\mathbb B}
\newcommand{\bP}{\mathbb P}
\newcommand{\bF}{\mathbb F}
\newcommand{\bS}{\mathbb S}
\newcommand{\cA}{\mathcal A}
\newcommand{\cB}{\mathcal B}
\newcommand{\cF}{\mathcal F}
\newcommand{\cG}{\mathcal G}
\newcommand{\cP}{\mathcal P}
\newcommand{\cM}{\mathcal M}
\newcommand{\cT}{\mathcal T}
\newcommand{\cV}{\mathcal V}
\newcommand{\cX}{\mathcal X}
\newcommand{\xbar}{\bar x}

\newcommand{\Dx}{\nabla_x}
\newcommand{\Dxx}{\nabla^2_x}

\newcommand{\norm}[1]{\lVert #1 \rVert}
\newcommand{\braket}[2]{\langle#1,#2\rangle}

\title{Optimal Transport for Model Calibration}

\author{Ivan Guo$^1$ $^2$}
\author{Gr\'egoire Loeper$^1$$^2$$^4$}
\author{Jan Ob\l\'oj$^3$}
\author{Shiyi Wang$^1$}
\address{$^1$School of Mathematics, Monash University, Clayton, VIC, Australia}
\address{$^2$Centre for Quantitative Finance and Investment Strategies,  Monash University, Clayton, VIC, Australia}
\address{$^3$University of Oxford, Oxford, United Kingdom}
\address{$^4$BNP Paribas Global Markets}

\begin{abstract}
We provide a survey of recent results on model calibration by Optimal Transport. We present the general framework and then discuss the calibration of local, and local-stochastic, volatility models to European options, the joint VIX/SPX calibration problem as well as calibration to some path-dependent options. We explain the numerical algorithms and present examples both on synthetic and market data. 
\end{abstract}

\maketitle

\section{Introduction}
In recent years, optimal transport theory has attracted the attention of many researchers. The problem was first formulated by Monge \cite{monge1781memoire} in the context of civil engineering and was later given a rigorous mathematical treatment by Kantorovich \cite{kantorovich1948} through the introduction of linear programming (for which he was awarded the Nobel Prize in economics in 1970). Brenier \cite{Br1} in 1991 then revisited the subject (having in mind applications to the famous Euler equations of fluid dynamics), and later on, in 2000, Benamou and Brenier \cite{benamou-brenier2000} introduced a time-continuous formulation of the problem, which gave rise to a massive amount of applications and mathematical results, see \cite{Vi, Vi2} for an account of these results. Two recent Fields medallists (Villani 2010 and Figalli 2018) are world specialists of optimal transport, which says a lot about the importance that the topic has taken nowadays.

Recently, the theory of optimal transport has been adapted to solve problems in robust hedging and pricing both in discrete and in continuous-time models, see \cite{BHLP:13,HenryLabordere:2014hta, tan-touzi2013, de2015linking}, when it was discovered that pricing bounds on path-dependent derivatives with fixed European options could be formulated as a \emph{martingale optimal transport} problem. Martingale optimal transport then became a subject of study in itself. The theory has been further used to calibrate the non-parametric discrete-time model proposed by Guyon \cite{guyon2020joint} (see \cite{guyon2021dispersion} for an extended version) to solve the so-called VIX/SPX calibration problem. The Schr\"odinger bridge problem, which is highly related to optimal transport, has been recently applied by Henry-Labod\`ere \cite{henry2019martingale} to introduce a new class of stochastic volatility models. These models can be calibrated by modifying only the drift while keeping the volatility of volatility unchanged.

In this paper, we give a synthetic overview of recent results on the continuous-time optimal transport for model calibration, obtained in \cite{guo2018path,guo2020joint,guo2019calibration,guo2017local}. The results allow for exact calibration in the spirit of the celebrated Dupire's formula, albeit without requiring the the knowledge of prices for a continuum of European options. We review the calibration of
\begin{itemize}
\item[-] local volatility models to European options \cite{guo2017local},
\item[-] local-stochastic volatility models to European options \cite{guo2019calibration}, 
\item[-] the joint VIX/SPX calibration problem \cite{guo2020joint},
\item[-] path-dependent models to path-dependent options (e.g., Asian, barrier and lookback options) \cite{guo2018path}. 
\end{itemize}
In particular, to the best of our knowledge, the last result was the first rigorous calibration framework including non-European options.

\section{The semimartingale optimal transport problem}\label{sec:sot}

\subsection{Probabilistic formulation}
 
The problem of optimal transport by semimartingales was studied by Tan and Touzi \cite{tan-touzi2013}. Later in \cite{guo2019calibration}, motivated by financial applications, the authors extended this problem by replacing the terminal distribution constraint with a finite number of discrete constraints. 

Let $\Omega:=C([0,T],\R^d), T>0$ be the set of continuous paths, $X$ be the canonical process and $\bF=(\cF_t)_{0\leq t\leq T}$ be the canonical filtration generated by $X$. Let $\cP^0$ be the collection of all probability measures $\bP$ on $(\Omega, \cF_T)$, under which $X$ is an $(\bF,\bP)$-semimartingale such that
\eqn{
  dX_t = \alpha_t^\bP\,dt + (\beta_t^\bP)^\demi\,dW_t^\bP,
}
where $W^\bP$ is a $\bP$-Brownian motion, and $(\alpha^\bP,\beta^\bP)$ are $\bF$-adapted processes. In particular, we say that $\bP$ is \emph{characterised} by $(\alpha^\bP,\beta^\bP)$. Let $\cP^1\subset\cP^0$ be a subset of probability measures $\bP$ characterised by $(\alpha^\bP, \beta^\bP)$ that are $\bP$-integrable on $[0,T]$, i.e.,
\eqn{
  \bE^\bP\left(\int_0^T |\alpha_t^\bP| + |\beta_t^\bP| \,dt \right) < +\infty,
}
where $|\cdot|$ is the Euclidean norm. 

$\cP^1$ corresponds to the set of feasible market dynamics. Throughout, for simplicity, we will take the interest rates and the dividend yield to be zero\footnote{In applications with market data we then work out the forward prices.}. To consider the subset of calibrated models, we fix $x_0\in\R^d$ and a finite number $m$ of constraints: market prices $c\in\R^m$ corresponding to options with payoffs $\cG\in(C_b(\R^d))^m$ and maturities $\cT\in(0,T]^m$.  We assume that the longest maturity coincides with the time horizon, $\max_k \cT_k=T$. We are then interested in:
\eqn{
  \cP(x_0, c, \cT, \cG):=\{ \bP\in\cP^1 : \bP\circ X_0^{-1}=\delta_{x_0} \mbox{ and } \bE^\bP \cG_i(X_{\cT_i})=c_i ,\, i=1,\ldots,m \}. 
} 

We may have further restrictions on the pricing measures, e.g., some assets may have to be martingales. This, as well as other desirable properties, e.g., proximity to a reference model, are encoded through a cost function $F:[0,T]\times\R^d\times\R^d\times\bS^d\to\R\cup\{+\infty\}$, where $\bS^d$ denotes the set of symmetric matrices of order $d$. $F$ is taken  convex in $(\alpha_t^\bP, \beta_t^\bP)$. Finding a suitable calibrated market model corresponds to solving 
\eq{\label{eq:prob_formulation}
  V:=\inf_{\bP\in\cP(x_0,c,\cT,\cG)} \bE^\bP\int_0^T F(t,X_t,\alpha_t^\bP, \beta_t^\bP)\,dt,
}
where $\inf\emptyset = +\infty$ and, in particular,  a finite value indicates that a perfectly calibrated model was found. 

\subsection{PDE formulation}

If in the above problem the state variables are matched to the constraints then the mimicking properties of diffusions, see \cite{brunick2013}, allow us to restrict the optimisations to \emph{local} diffusions, i.e., to $(\alpha^\bP,\beta^\bP)$ which are functions of time $t$ and the state variables $X_t$. Therefore, the problem in \eqref{eq:prob_formulation} can be studied via PDE methods. Following the Benamou--Brenier formulation of the classical optimal transport from \cite{benamou-brenier2000}, we introduce the following formulation:
\begin{formulation}[PDE formulation]
Solve
\eqn{
  V=\inf_{\rho,\alpha,\beta}\int_0^T\int_{\R^d} F(t,x,\alpha(t,x),\beta(t,x))\rho(t,x)\,dxdt,
}
where the infimum is taken among all $(\rho,\alpha,\beta)$ satisfying (in the distributional sense)
\eqn{
  \dt\rho(t,x) + \sum_i \partial_i (\rho(t,x)\alpha_i(t,x)) - \demi\sum_{i,j}\partial_{ij}(\rho(t,x)\beta_{ij}(t,x))&=0,\\
  \int_{\R^d}\cG_i(x)\rho(\cT_i,x)\,dx &= c_i, \quad \forall i = 1,\ldots,m,\\
  \rho(0,\cdot) &= \delta_{x_0}.
}
\end{formulation}

\subsection{Dual formulation}

In the PDE formulation, the objective function is convex and all constraints are linear in $(\rho,\rho\alpha,\rho\beta)$. Applying the classical tools of convex analysis\footnote{The proof of duality mainly relies on the Fenchel--Rockafellar theorem. We refer the reader to \cite{guo2019calibration} for the full proof.}, we introduce a dual formulation:
\begin{formulation}[Dual formulation]
Solve
\eqn{
  V = \sup_{\lambda\in\R^m} \lambda\cdot c - \phi(0,x_0),
}  
where $\phi$ solves the following HJB equation (in the viscosity sense\footnote{See \cite{guo2019calibration} for the definition of the viscosity solution to \eqref{eq:generic_hjb}.}):
\eq{\label{eq:generic_hjb}
\dt\phi(t,x) + F^*(t,x,\Dx\phi(t,x),\demi\Dxx\phi(t,x)) = -\sum_{i=1}^m\lambda_i\cG_i(x)\delta(t-\cT_i)
}
with the terminal condition $\phi(T,\cdot)=0$, where $F^*$ is the convex conjugate of $F$ defined by
\eqn{
  F^*(t,x,a,b) = \sup_{\alpha,\beta}\{ \alpha\cdot a + \demi\sum_{ij}\beta_{ij}b_{ij} - F(t,x,\alpha,\beta)  \}.
}
\end{formulation}

If the optimal $\lambda$ has been found, one can obtain the optimal $(\alpha,\beta)$ of the PDE formulation by solving the supremum of $F^*$ in \eqref{eq:generic_hjb}. 

The dual formulation can be solved by gradient descent methods, and each component of the gradient vector can be calculated by solving a linear PDE. Given a $\lambda\in\R^m$, denote by $\phi^\lambda$ the associated solution to \eqref{eq:generic_hjb}. Let $(\alpha^\lambda,\beta^\lambda)$ be the maximisers in the definition \eqref{eq:generic_hjb} of $F^*$ with $\phi^\lambda$. Define $L(\lambda):=\sum_{i=1}^m\lambda_i c_i - \phi^\lambda(0,x_0)$, then $\partial_{\lambda_i}L(\lambda) = c_i - \partial_{\lambda_i}\phi^\lambda(0,x_0)$. Since $\phi^\lambda$ depends on $\lambda$, by taking functional derivatives of \eqref{eq:generic_hjb} with respect to $\lambda_i$, we can formulate the gradients as 
\eq{\label{eq:generic_gradient}
  \partial_{\lambda_i}L(\lambda) = c_i - \phi_i'(0,x_0), \quad i=1,\ldots,m,
}
where $\phi_i'$ solves
\eq{\label{eq:generic_pricing_PDE}
\left\{\begin{array}{l}
  \displaystyle \dt\phi'_i + \alpha^\lambda\cdot\Dx\phi'_i + \demi\sum_{ij}\beta^\lambda_{ij}\partial_{ij}\phi'_i = 0, \qquad \mbox{in } [0,\cT_i)\times\R^d, \\
  \displaystyle \phi'_i(\cT_i,\cdot) = \cG_i.
\end{array} \right.
}
Since $\phi'_i(0,x_0)=\bE^{\bP(\lambda)}\cG_i(X_{\cT_i})$ where $\bP(\lambda)$ is characterised by $(\alpha^\lambda,\beta^\lambda)$, the gradient $\partial_{\lambda_i}L(\lambda) = c_i - \bE^{\bP(\lambda)}\cG_i(X_{\cT_i})$ that can be interpreted as the difference between the option prices given by the current optimisation iteration (or simply model prices) and the market option prices. The optimum is reached when  the gradient is zero, in other words, the market option prices are attained exactly. 

\begin{remark}
  The problem of allowing $F$ and $\cG$ to be path-dependent was studied in \cite{guo2018path}. In that case, we need to solve path-dependent PDEs instead of HJB equations. We will show an example of calibrating a path-dependent volatility model to barrier options in Section \ref{sec:path-dependent} below.
\end{remark}

\begin{remark}
Recall that $\cG$ are required to be bounded continuous functions due to technical reasons. In practice, many options do not have bounded payoffs (e.g., call options). This can be fixed by either converting them into options with bounded payoffs via arbitrage arguments (e.g., put options via put-call parity), or by truncating the domain at some extremely large value. Options that do not have continuous payoffs (e.g., digital options, barrier options, etc.) can be approximated by uniformly continuous functions.
\end{remark}

\subsection{Numerical method}\label{sec:numerical_method}

A numerical method for solving the dual formulation was proposed in \cite{guo2019calibration}. The method can be described as follows:
\begin{enumerate}[label=(\roman*)]
  \item set an initial $\lambda$, e.g., $\lambda=\mathbf{0}\in\R^m$,
  \item obtain $\phi^\lambda(0,x_0)$ by solving the backward HJB equation and obtain $(\alpha^\lambda,\beta^\lambda)$ by solving the supremum of $F^*$ in \eqref{eq:generic_hjb},
  \item solve the linear pricing PDEs \eqref{eq:generic_pricing_PDE} with $(\alpha^\lambda,\beta^\lambda)$, and then calculate the gradients by \eqref{eq:generic_gradient},
  \item update $\lambda$ by a gradient descent algorithm,
  \item repeat step (ii)-(iv) until the all components of gradients are close to zero.
\end{enumerate}

In \cite{guo2019calibration} and \cite{guo2020joint}, the HJB equation \eqref{eq:generic_hjb} was solved by the standard implicit finite difference method with the so-called policy iteration technique to handle the nonlinearity, and the linear pricing PDEs \eqref{eq:generic_pricing_PDE} were solved by an alternating direction implicit finite difference method that is faster than the standard implicit method. For the gradient descent algorithm, the L-BFGS algorithm was employed and showed good convergence in both works. 

It should be mentioned that the dual formulation and the numerical method were also studied in \cite{avellaneda1997calibrating} much earlier in the context of uncertain volatility calibration via entropy minimisation, although the connection to optimal transport and the proof the duality result was not established at that time. 

\section{Applications in model calibration}

From now on, we will refer to the proposed calibration method simply as \emph{OT framework}.

\subsection{Local volatility calibration}\label{sec:local_vol}

The application of optimal transport to calibrate the local volatility model of Dupire \cite{Dupire1994pricing} was explored in \cite{guo2017local}. In \cite{guo2017local}, as an extension of the seminal work of \cite{benamou-brenier2000}, an augmented Lagrangian method was developed to solve the PDE formulation. In this section, we resolve the local volatility calibration problem by the OT framework.

Let $X_t$ be the logarithm of the underlying stock price at time $t$. We are interested in finding a probability measure $\bP\in\cP^1$ with characteristics $(-\demi\sigma^2, \sigma^2)$ where $\sigma$ is some adapted process. In other words, we want $X$ to be a $\bP$-semimartingale in the form of 
\eq{\label{eq:dynamics_lv}
  dX_t=-\demi\sigma_t^2\,dt + \sigma_t\,dW_t^\bP.
}
To ensure that $X$ solves the above SDE, we consider a cost function of the form
\eq{\label{eq:costfun_lv}
  F(t,x, \alpha,\beta) = \left\{\begin{array}{ll}
  a(\beta/\bar\sigma^2)^p + b(\beta/\bar\sigma^2)^{-q} + c,  &-2\alpha = \beta>0,\\
  +\infty, & \mbox{otherwise},
\end{array} \right.
}
where $\bar\sigma$ is some reference volatility level, $p,q$ are constants greater than 1, and $a, b, c$ are constants chosen so that the function reaches its minimum at $\beta=\bar\sigma^2$ with $\min F = 0$.    

Given a vector of $m$ (discounted) European option payoff functions $G$, a vector of maturities $\tau$ and a vector of European option prices $c$, we want to further restrict $\bP$ so that $\bE^\bP G_i(X_{\tau_i})=c_i,\, i=1,\ldots,m$ are satisfied. Let $x_0$ be the logarithm of the current stock price, then the local volatility calibration problem can be reformulated as solving
\eq{\label{eq:main_lv_calibration}
  V_{LV} := \inf_{\bP\in\cP(x_0, c, \tau, G)}\bE^\bP\int_0^T F(t,X_t,\alpha^\bP_t,\beta^\bP_t)\,dt,
}
where
\eqn{
  \cP(x_0, c, \tau, G):=\{ \bP\in\cP^1 : \bP\circ X_0^{-1}=\delta_{x_0} \mbox{ and } \bE^\bP G_i(X_{\tau_i})=c_i ,\, i=1,\ldots,m \}.
} 

Following the OT framework, we introduce the dual formulation of \eqref{eq:main_lv_calibration}:
\eq{\label{eq:dual_objective_lv}
  V_{LV} = \sup_{\lambda\in\R^m} \lambda\cdot c - \phi(0,x_0),
}
where $\phi$ is a solution to the HJB equation (in the viscosity sense)
\eq{\label{eq:hjb_lv}
  \dt\phi + \sup_{\beta>0}\left\{ -\demi\beta\dx\phi + \demi\beta\dxx\phi
  - \left(a\left(\frac{\beta}{\bar\sigma^2}\right)^p + b\left(\frac{\beta}{\bar\sigma^2}\right)^{-q} + c\right)\right\} = -\sum_{i=1}^m\lambda_i G_i\delta(t-\tau_i),
}
with the terminal condition $\phi(T,\cdot) = 0$.

\subsubsection*{Numerical example}

\begin{figure}[ht!]
\begin{center}
\includegraphics[width=1.0\textwidth]{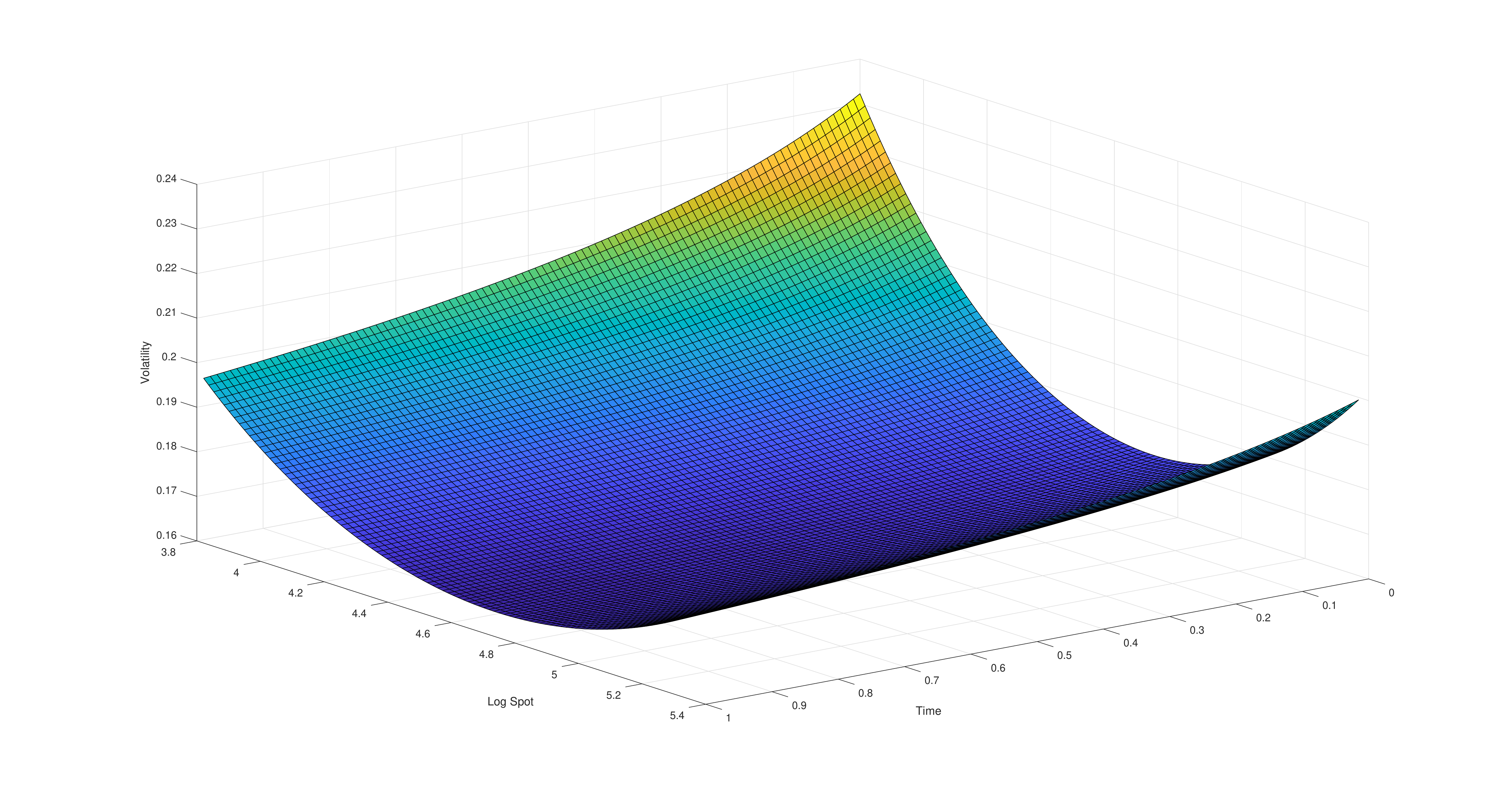}
\caption{The local volatility surface used for generating European put option prices.}
\label{fig:given_local_vol}
\end{center}
\end{figure}

We give here an example in which a local volatility model is calibrated to the prices of 5 European put options at 5 different strikes and maturity $T=1$. The option prices are generated by another local volatility model with the volatility given in Figure \ref{fig:given_local_vol}. The value of $\bar\sigma$ in \eqref{eq:costfun_lv} is set to 0.2.

\begin{figure}[ht!]
  \begin{minipage}{0.5\textwidth}
    \centering
    \includegraphics[width=1.0\linewidth]{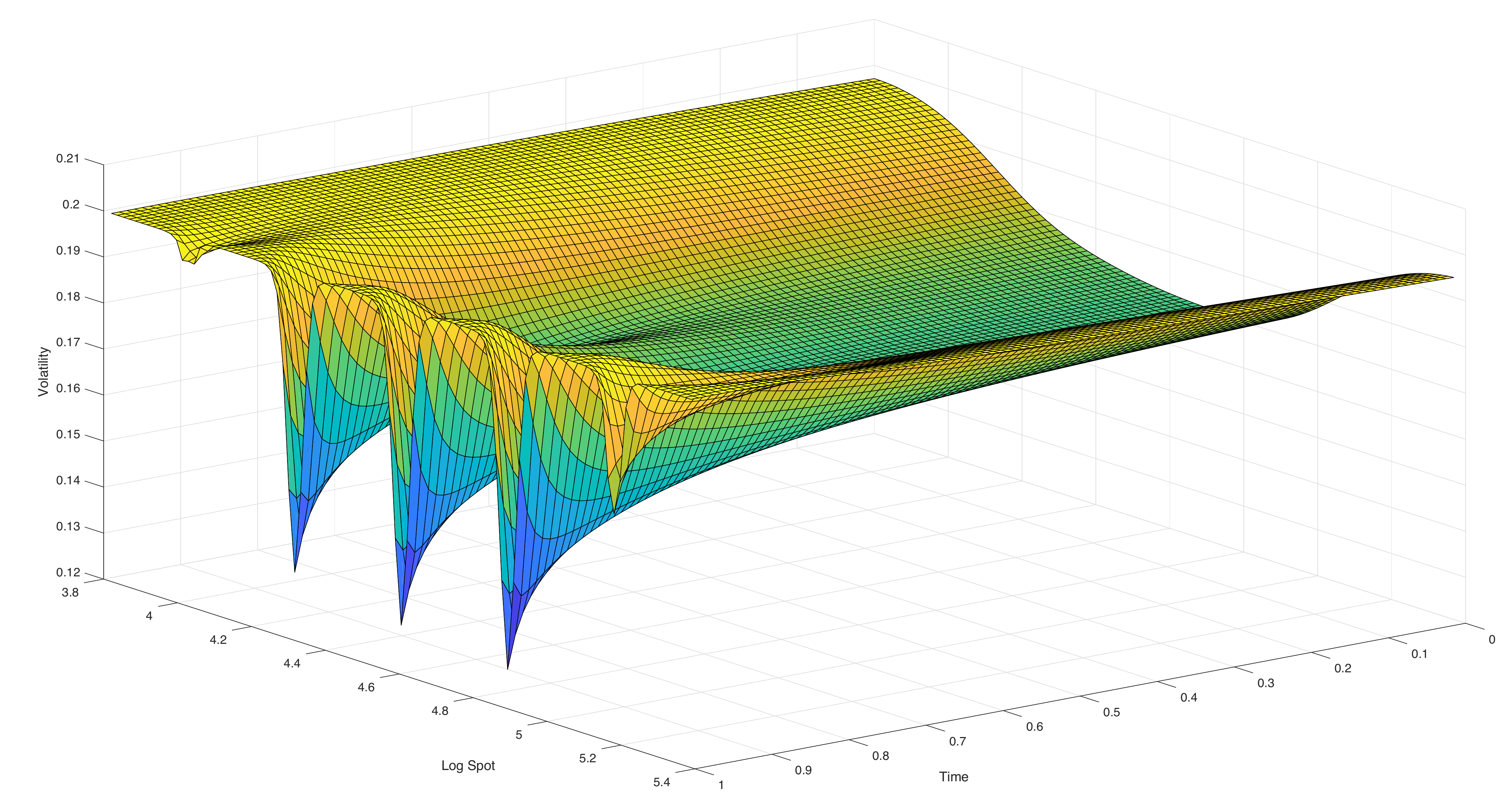}
  \end{minipage}\hfill
  \begin{minipage}{0.5\textwidth}
    \centering
    \includegraphics[width=1.0\linewidth]{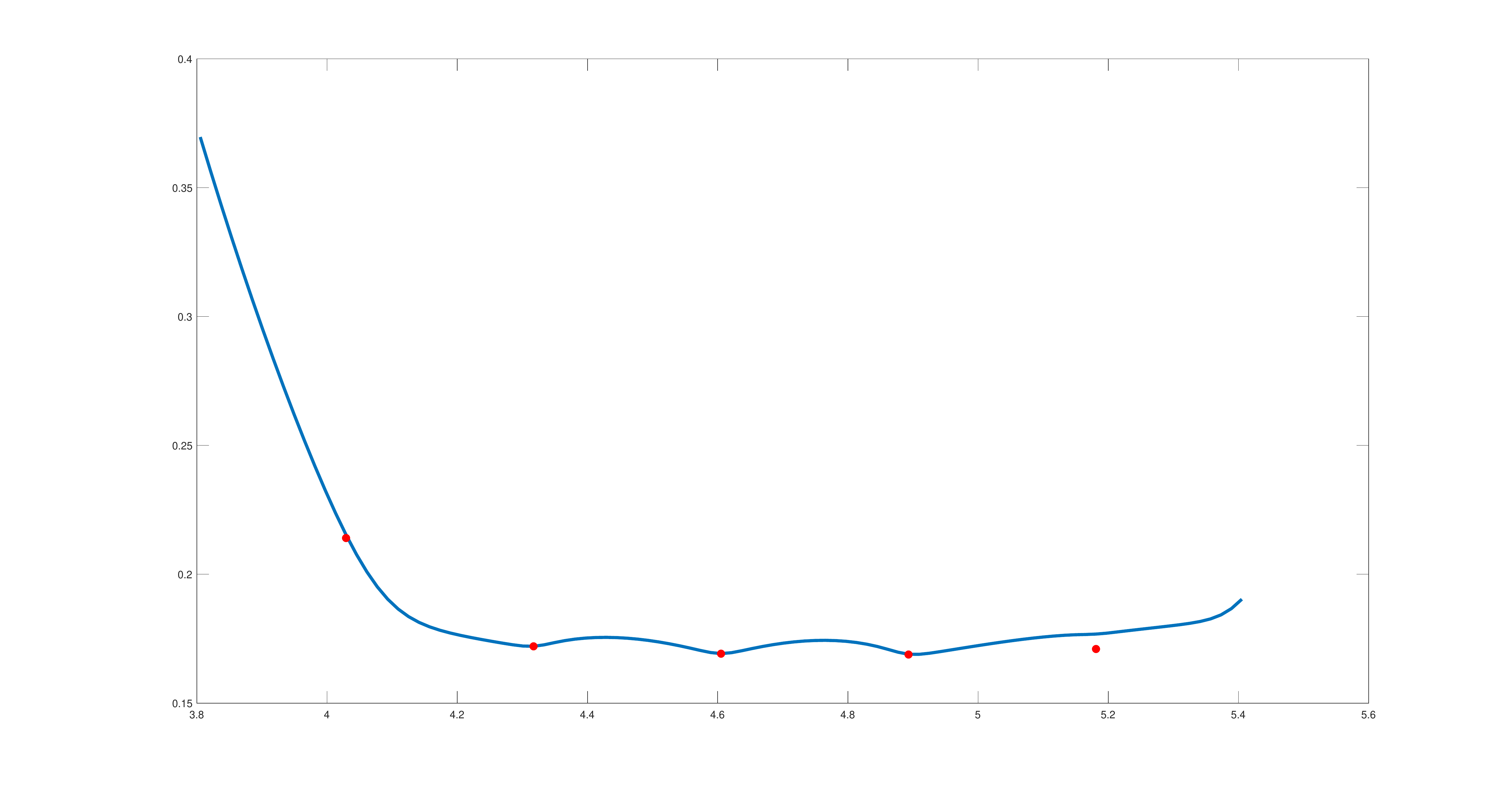}
  \end{minipage}
  \caption{The (unsmoothed) calibrated local volatility surface (left), the model volatility skew (right, blue) and the implied volatility of the calibrating options (right, red).}
  \label{fig:calibrated_local_vol}
\end{figure}

Figure \ref{fig:calibrated_local_vol} shows the calibrated local volatility surface and the model implied volatility skew. The humps between strikes in the volatility skew are caused by the spikes in the volatility surface. These spikes were also observed in \cite{avellaneda1997calibrating}. 

\begin{figure}[ht!]
  \begin{minipage}{0.5\textwidth}
    \centering
    \includegraphics[width=1.0\linewidth]{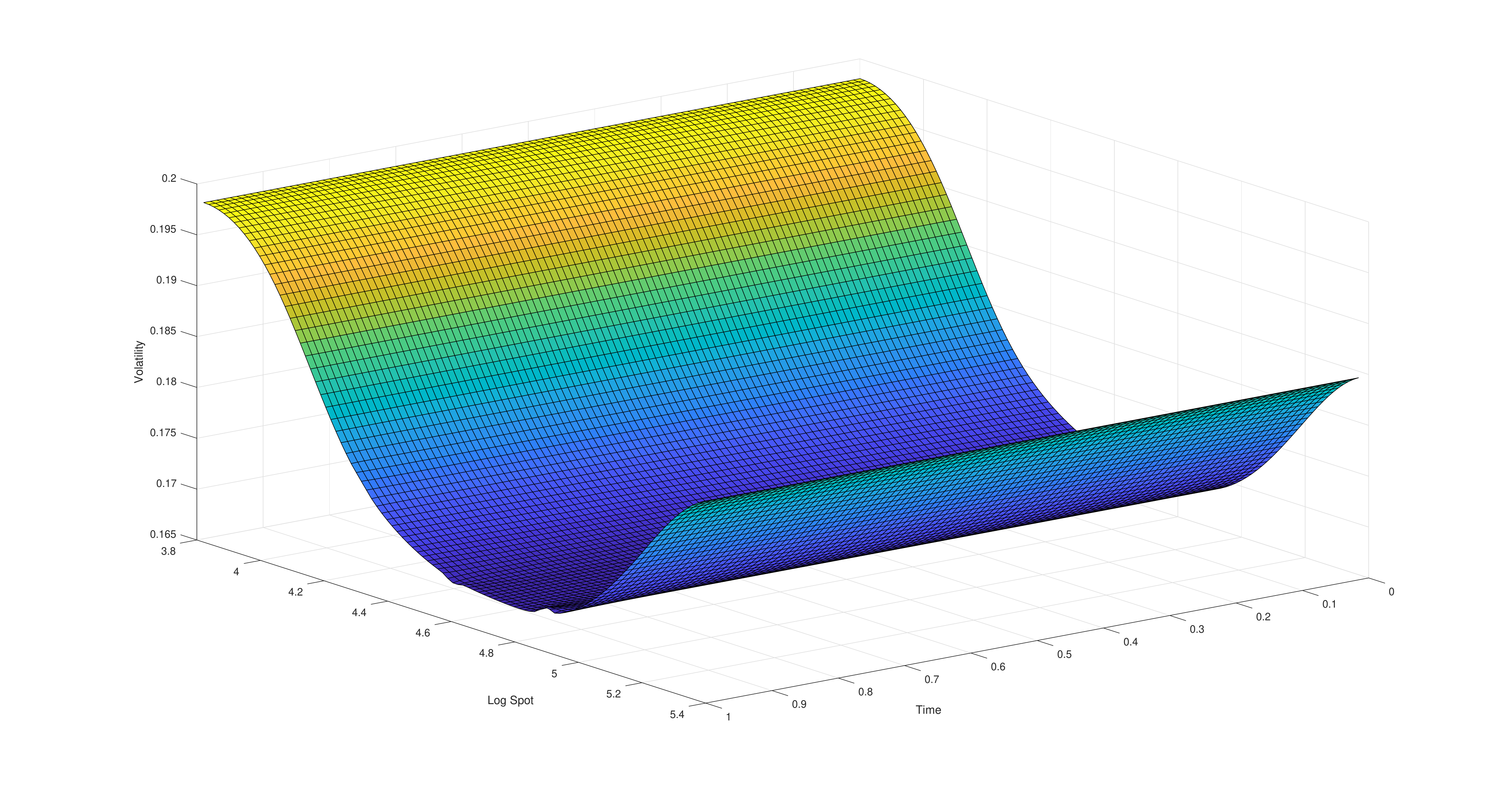}
  \end{minipage}\hfill
  \begin{minipage}{0.5\textwidth}
    \centering
    \includegraphics[width=1.0\linewidth]{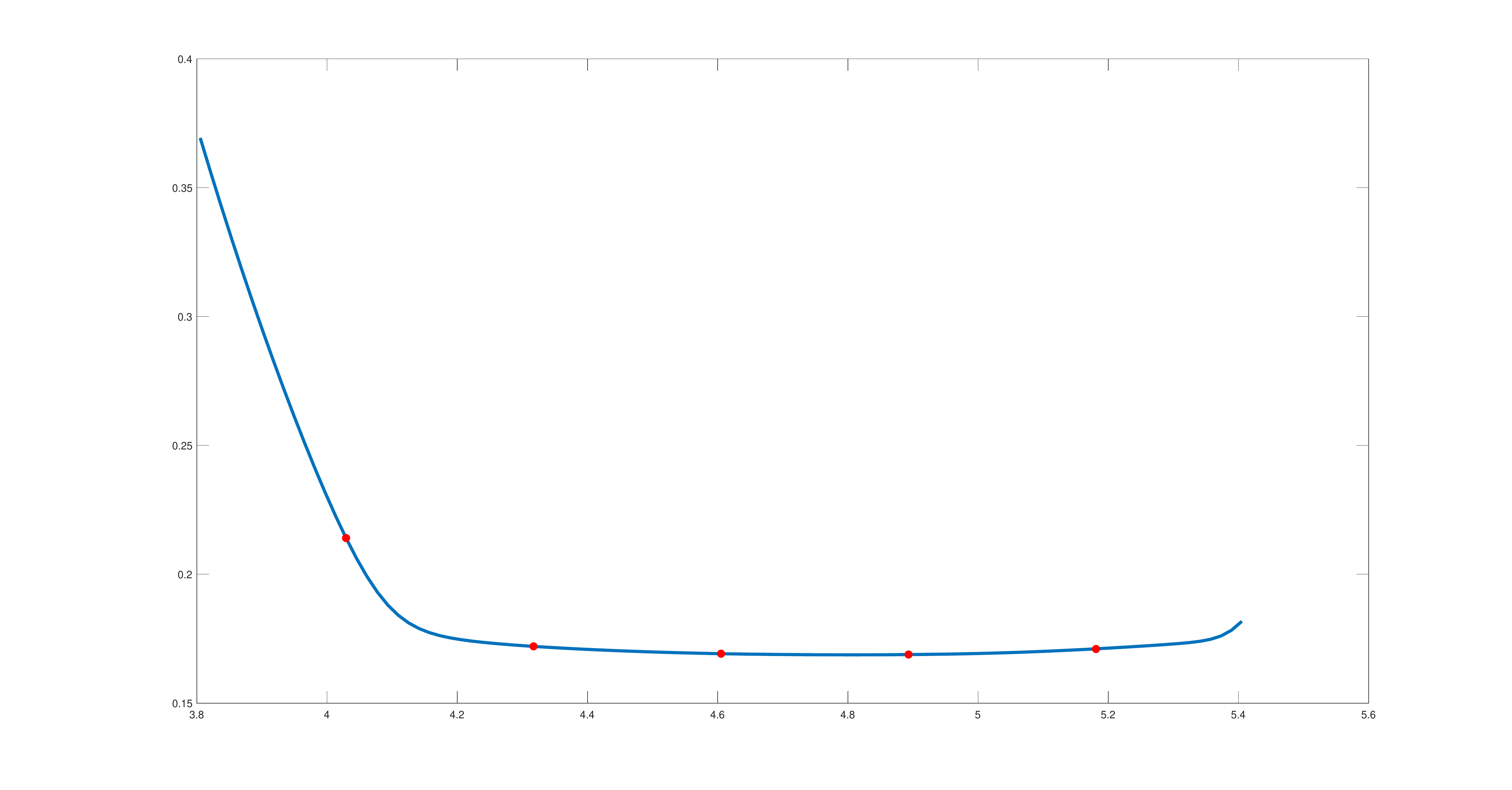}
  \end{minipage}
  \caption{The smoothed local volatility surface (left), the model volatility skew after 8 iterations (right, blue) and the implied volatility of the calibrating options (right, red).}
  \label{fig:smoothed_local_vol}
\end{figure}

To smooth the volatility surface and hence the volatility skew, we suggest a \emph{reference iteration} method. We start smoothing the spiky volatility surface by a simple moving average method. Next, we set the smoothed surface as the reference value $\bar\sigma$ and recalibrate the model by solving the dual formulation \eqref{eq:dual_objective_lv}. After iterating the above steps 8 times, we obtain a local volatility model that has a smooth volatility skew and is also fully calibrated to the given option prices. The results are shown in Figure \ref{fig:smoothed_local_vol}.

\subsection{Local Stochastic Volatility Calibration}

The Local-stochastic volatility (LSV) model was first introduced in \cite{jex1999}. It incorporates a nonparametric local factor (also called \emph{leverage}) into a classical stochastic volatility model. Thus, while keeping consistent dynamics, the LSV model can match all observed market option prices, as long as one restricts to European products. In this section, we apply the OT framework to solve the LSV model calibration problem, as done in \cite{guo2019calibration}.

Consider probability measures $\bP\in\cP^1$ under which $X=(X^1, X^2)$ are two-dimensional $\bP$-semimartingales. Let $X^1$ be the logarithm of the underlying price and let $X^2$ be a mean-reverting stochastic factor. We are particularly interested in the following LSV model
\eq{\label{eq:dynamics_lsv}
  \begin{cases}
    dX^1_t = -\demi\sigma_t^2\,dt + \sigma_t \,dW_t^1,\\
    dX^2_t = \kappa(\theta - X^2_t)\,dt + \xi\sqrt{X^2_t}\,dW_t^2,\\
    \displaystyle dW_t^1 dW_t^2 = \eta\frac{\sqrt{X^2_t}}{\sigma_t} \,dt,
  \end{cases}
}
where $\sigma$ is some adapted process, and $(\kappa,\theta,\xi,\eta)$ are constant parameters and are assumed given. The above model dynamics can be captured by probability measures $\bP$ characterised by $(\alpha^\bP,\beta^\bP)$ such that
\eqn{
  (\alpha^\bP_t,\beta^\bP_t) = \left( 
    \begin{bmatrix} 
      -\demi\sigma^2_t \\
      \kappa(\theta - X^2_t)
    \end{bmatrix},
    \begin{bmatrix}
      \sigma^2_t & \eta\xi X^2_t \\
      \eta\xi X^2_t & \xi^2 X^2_t
    \end{bmatrix}
  \right), \quad 0\leq t\leq T.
}
The model we consider here is slightly different from the standard LSV model from the literature. In the standard LSV model, the correlation between $W^1$ and $W^2$ is a constant and $\sigma_t = L(t,X^1_t)\sqrt{X^2_t}$, where $L$ is known as the leverage function. Our simple modification allows that if a function is convex in $\beta$, then it is convex in $\sigma^2$, which makes it easier to define a suitable cost function. Note that if $\sigma_t=\sqrt{X^2_t}$, the correlation is simply $\eta$ and $X$ reduces to a Heston model. If we define a cost function to penalise $\sigma_t$ away from $\sqrt{X^2_t}$, and we obtain $(\kappa,\theta,\xi,\eta)$ by calibrating a Heston model to the market prices, then $\sigma_t$ will be close to $\sqrt{X^2_t}$ and hence the correlation will be close to $\eta$. Moreover, if $\sigma_t$ is independent of $X^2_t$, then $X$ is indeed a local volatility model, and $X$ can be exactly calibrated to the option prices generated by any arbitrage-free implied volatility surface. Our goal is to calibrate $\sigma_t$ with given $(\kappa,\theta,\xi,\eta)$ so that $X$ is fully calibrated to the observable market European option prices. 

In the spirit of \eqref{eq:costfun_lv}, let us first define a convex function
\eqn{
  H(x,\xbar,s):= \left\{ \begin{array}{ll}\displaystyle a\left(\frac{x-s}{\xbar-s}\right)^{p} + b\left(\frac{x-s}{\xbar-s}\right)^{-q} + c  & \mbox{if } x > s \mbox{ and } \xbar > s, \\
		   +\infty & \mbox{otherwise,} \end{array} \right.
}
where $p, q$ are constants greater than 1, and $a, b, c$ are constants chosen so that the function reaches its minimum at $x=\xbar > s$ with $\min H = 0$. To ensure that $X$ has the dynamics \eqref{eq:dynamics_lsv}, we define the cost function
\eqn{
  F(t,x,\alpha,\beta) = \left\{ \begin{array}{ll} H(\beta_{11}, x_2, \eta^2 x_2)  & \mbox{if } (\alpha,\beta)\in \Gamma(t,x), \\
		   +\infty & \mbox{otherwise,} \end{array} \right.
}
where the convex set
\eqn{
  \Gamma(t,x):=\{ (\alpha,\beta) \mid \alpha_1=-\beta_{11}/2, \alpha_2=\kappa(\theta-x_2), \beta_{12}=\beta_{21}=\eta\xi x_2, \beta_{22}=\xi^2 x_2  \}.
}
In the function $H$, we set $s=\eta^2 x_2$ to keep $\sigma^2_t >\eta^2 X^2_t$ so that $\beta_t$ remains positive semidefinite for all $t\leq T$. We set $\xbar = x_2$ to regularise $\beta_{11}$ (or $\sigma^2$) by penalising deviations of $X$ from a standard Heston model. 

Given a vector of $m$ (discounted) European option payoff functions $G$\footnote{Note that $G$ are functions of $X$. For example, if $G_i$ is the payoff function of an European call option, $G_i(x)=\max(\exp(x_1)-K, 0), K>0$.}, a vector of maturities $\tau$ and a vector of European option prices $c$, we want to further restrict $\bP$ so that $\bE^\bP G_i(X_{\tau_i})=c_i,\, i=1,\ldots,m$ are satisfied. Assume that $x_0\in\R^2$ is given. Its first element is the logarithm of the current stock price, which is observed from the market, and its second element is the initial value of the instantaneous variance, which is a parameter but can be obtained by calibrating a Heston model. Then the LSV model calibration problem can be reformulated as solving
\eq{\label{eq:main_lsv_calibration}
  V_{LSV} := \inf_{\bP\in\cP(x_0, c, \tau, G)}\bE^\bP\int_0^T F(t,X_t,\alpha^\bP_t,\beta^\bP_t)\,dt,
}
where
\eqn{
  \cP(x_0, c, \tau, G):=\{ \bP\in\cP^1 : \bP\circ X_0^{-1}=\delta_{x_0} \mbox{ and } \bE^\bP G_i(X_{\tau_i})=c_i ,\, i=1,\ldots,m \}.
} 
Applying the arguments developed in Section 2, we can derive a dual formulation of \eqref{eq:main_lsv_calibration}:
\eqn{
  V_{LSV} = \sup_{\lambda\in\R^m}\lambda\cdot c - \phi(0,x_0),
}
where $\phi$ is a solution to the following HJB equation (in the viscosity sense):
\eqn{
  \dt\phi + \sup_{\beta_{11}>0}\bigg\{ &-\demi\beta_{11}\partial_{x_1}\phi +\kappa(\theta-x_2)\partial_{x_2}\phi + \demi\beta_{11}\partial_{x_1x_1}\phi + \demi\xi^2x_2\partial_{x_2x_2}\phi \\
  & + \eta\xi x_2\partial_{x_1x_2}\phi -H(\beta_{11},x_2,\eta^2x_2)\bigg\} = -\sum_{i=1}^m\lambda_i G_i\delta(t-\tau_i),
}
with the terminal condition $\phi(T,\cdot) = 0$.

\subsubsection*{Numerical example}

In the numerical example provided in \cite{guo2019calibration}, the process $X$ in \eqref{eq:dynamics_lsv}, also called the \emph{OT-LSV} model, was calibrated to the FX options market data provided in \cite{tian2015calibrating}. The data contains 10 maturities ranging from 1 month to 5 years. At each maturity, there are 5 European options at different strikes. The parameters are $(\kappa,\theta,\xi,\eta) = (0.8721, 0.0276, 0.5338, -0.3566)$ which are obtained by (roughly) calibrating a standard Heston model to the market option prices. Since $2\kappa\theta/\xi^2=0.168\ll 1$, the Feller condition is strongly violated in this case. The initial position $X_0 =(0.2287, 0.012)$. 

\begin{figure}[ht!]
  \begin{minipage}{0.49\textwidth}
    \centering
    \includegraphics[width=1.0\linewidth]{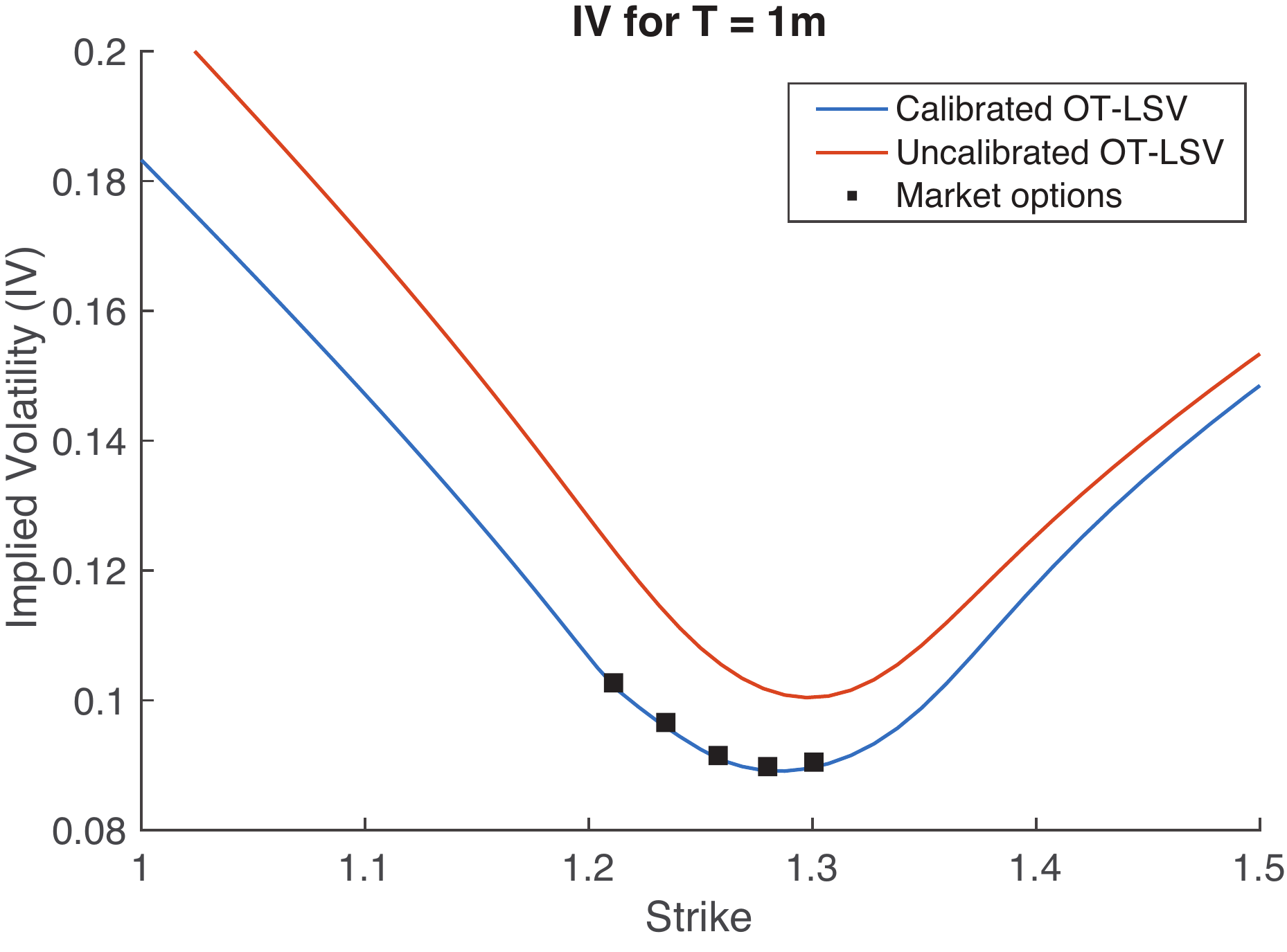}
  \end{minipage}
  \begin{minipage}{0.49\textwidth}
    \centering
    \includegraphics[width=1.0\linewidth]{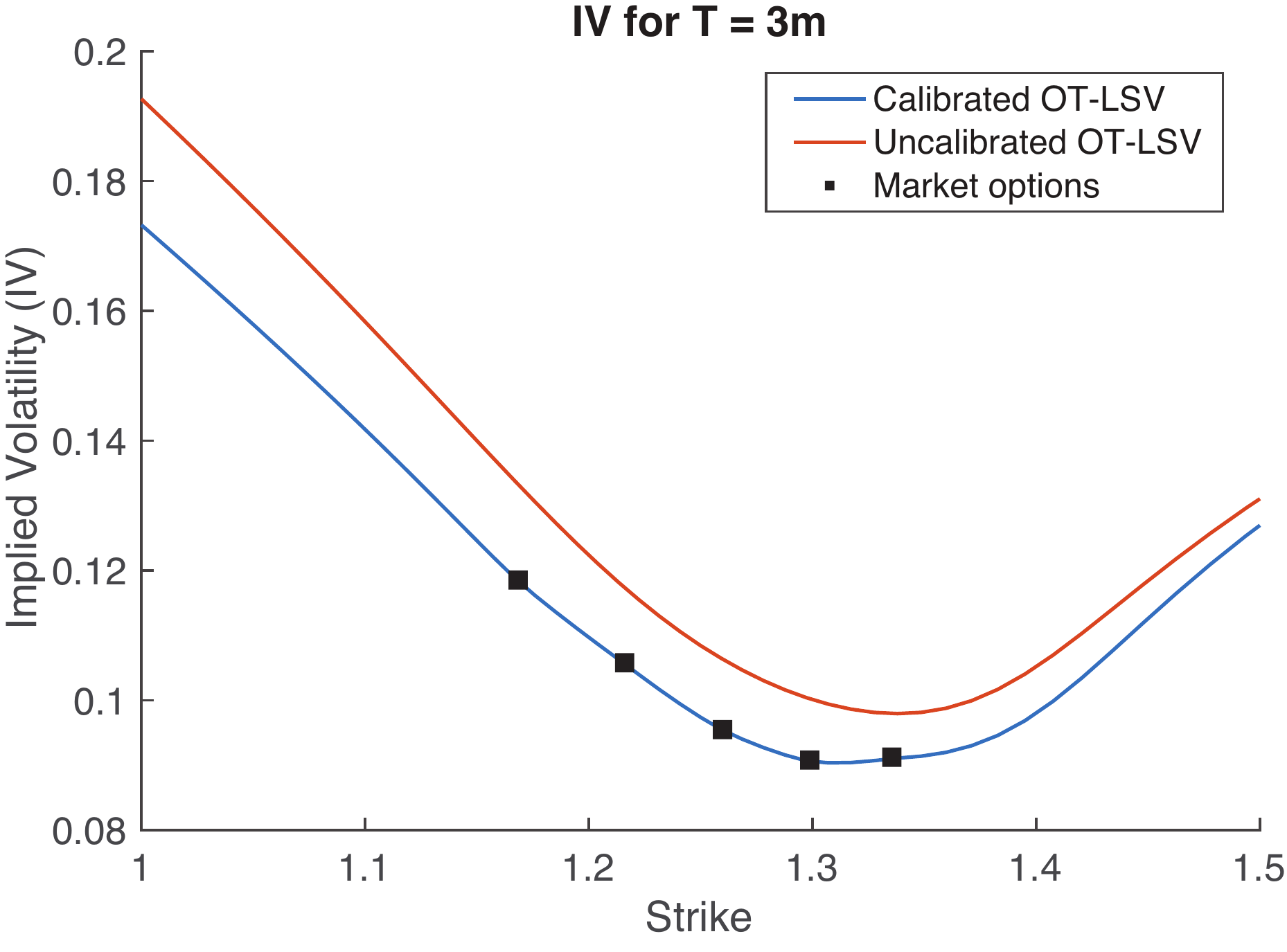}
  \end{minipage}
  \caption{The implied volatility skews generated by both the uncalibrated and the calibrated OT-LSV model for 1 month and 3 months maturities in the FX market data example.}
  \label{fig:LSV_IV_short}
\end{figure}

\begin{figure}[ht!]
  \begin{minipage}{0.49\textwidth}
    \centering
    \includegraphics[width=1.0\linewidth]{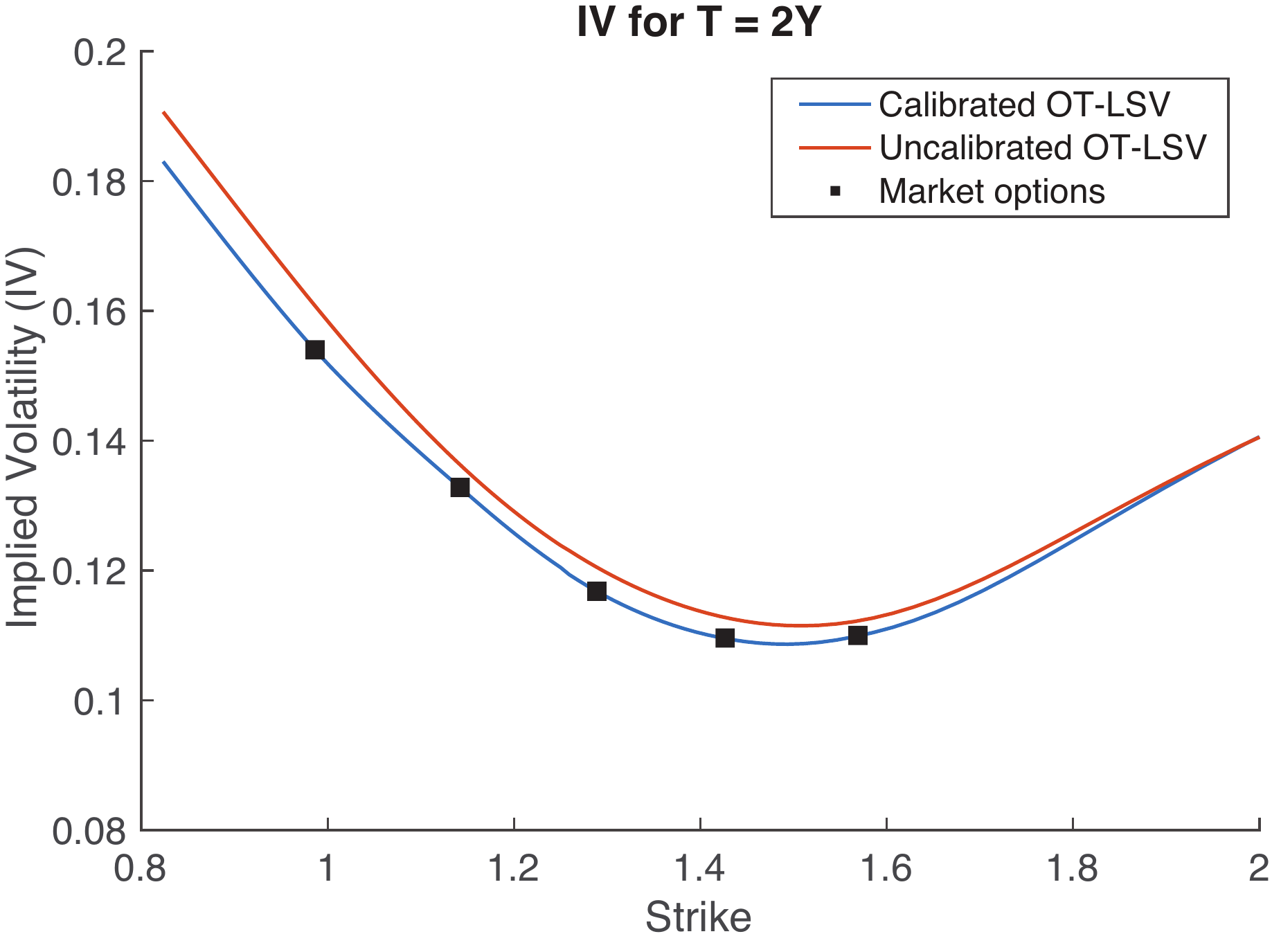}
  \end{minipage}\hfill
  \begin{minipage}{0.49\textwidth}
    \centering
    \includegraphics[width=1.0\linewidth]{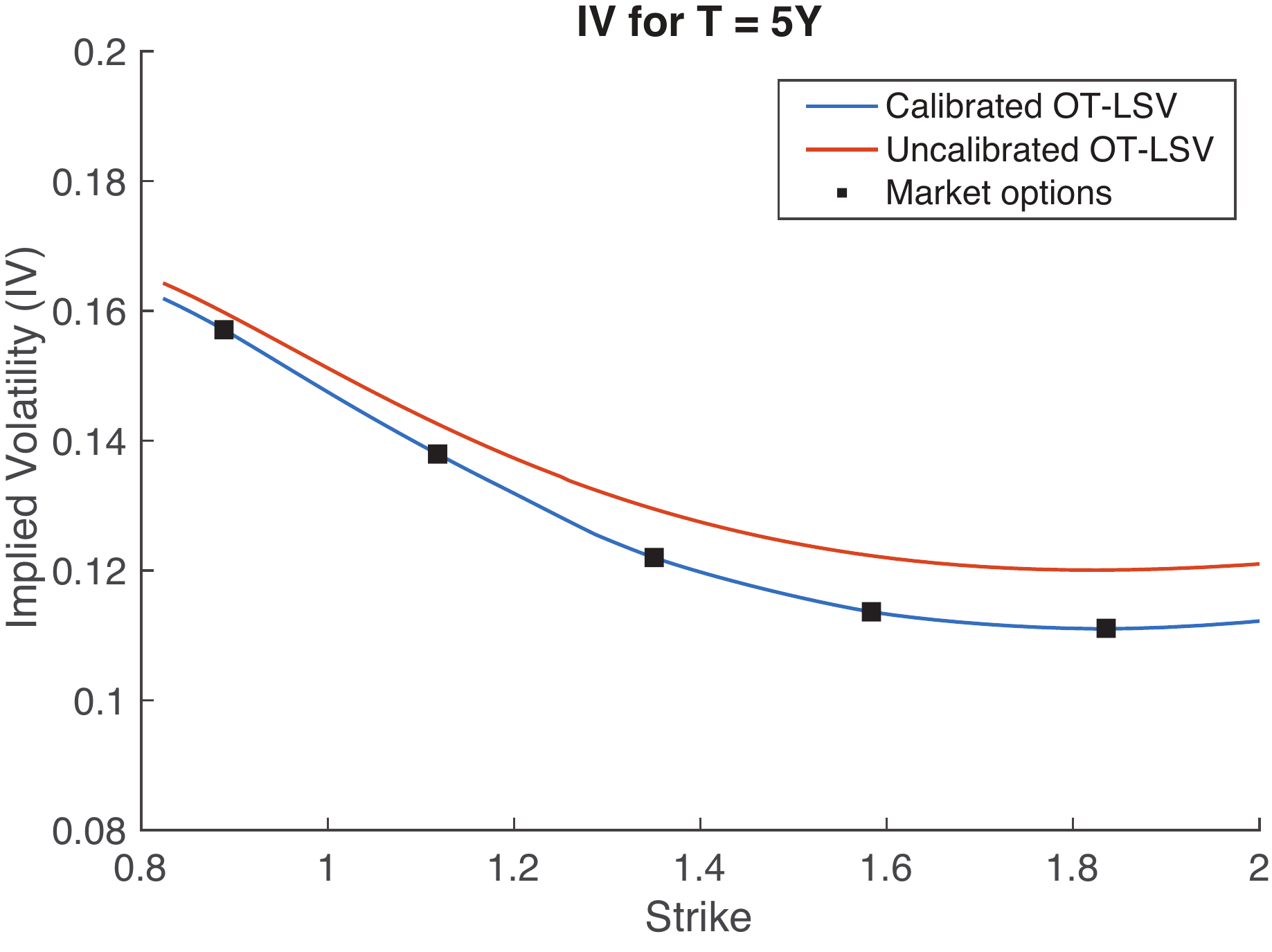}
  \end{minipage}
  \caption{The implied volatility skews generated by both the uncalibrated and the calibrated OT-LSV model for 2 years and 5 years maturities.}
  \label{fig:LSV_IV_long}
\end{figure}

Figures \ref{fig:LSV_IV_short} and \ref{fig:LSV_IV_long} compare the implied volatility skews of both the calibrated and uncalibrated OT-LSV model. The results show that the OT-LSV model can be accurately calibrated to both short-maturity and long-maturity market option prices. Unlike the local volatility example in Section \ref{sec:local_vol}, the volatility skews are very smooth even without iterating the reference values. 

\subsection{VIX/SPX joint calibration}

Since it was first reported in \cite{gatheral2008consistent}, the joint calibration on SPX and VIX has been known to be a challenging problem. More specifically, we want to build a stochastic volatility model that could be jointly calibrated to the options and futures of SPX and VIX. We refer the reader to \cite{guyon2020joint} for a comprehensive discussion of the literature and a martingale optimal transport approach with a discrete-time model. In this paper, we introduce the work of \cite{guo2020joint} in which the OT framework was applied to solve the joint calibration problem.

Consider probability measures $\bP\in\cP^1$ under which $X=(X^1, X^2)$ are two-dimensional $\bP$-semimartingales. We want $X^1$ to be the logarithm of the SPX price that takes the form of
\eq{\label{eq:dynamics_joint_x1}
  X^1_t=X^1_0 - \demi\int_0^t \sigma_s^2\,ds + \int_0^t\sigma_s\,dW_s, \quad 0\leq t \leq T.
}
For such $X^1$, we then use $X^2_t$ (or $X^2_{t,T}$ when emphasising the dependence on $T$) to represent a half of the expectation of the forward quadratic variation of $X^1$ on $[t,T]$ observed at time $t$, that is
\eq{\label{eq:dynamics_joint_x2}
  X^2_{t,T} = \bE^\bP\left(\demi\int_t^T\sigma^2_s\,ds \,\bigg|\, \cF_t\right) = X^1_t - \bE^\bP(X^1_T \mid \cF_t),\quad 0\leq t\leq T.
}
Note that the the second term on the right-hand side of (\ref{eq:dynamics_joint_x2}) is a martingale. It follows that the modelling setting we just described is captured by probability measures $\bP\in\cP^1$ characterised by $(\alpha,\beta)$ such that
\eq{\label{eq:characteristics_joint}
  (\alpha_t, \beta_t) = \left(
    \begin{bmatrix}
      -\demi\sigma_t^2 \\
      -\demi\sigma_t^2
    \end{bmatrix},
    \begin{bmatrix}
      \sigma_t^2 & (\beta_t)_{12} \\
      (\beta_t)_{12} & (\beta_t)_{22} 
    \end{bmatrix}
  \right), \quad 0\leq t\leq T,
}
where $(\beta_t)_{12} = d\langle X^1,X^2\rangle_t \mathbin{/} dt$ and $(\beta_t)_{22} = d\langle X^2\rangle_t \mathbin{/} dt$ and with the additional property that $X^2_{T,T}=0$ $\bP$-a.s. 

In order to restrict the probability measures to those characterised by $(\alpha,\beta)$ of the form (\ref{eq:characteristics_joint}), we can define a cost function that penalises characteristics that are not in the following convex set:
\eqn{
  \Gamma:= \left\{(\alpha,\beta)\in\R^2\times\bS^2_+ : \alpha_1=\alpha_2=-\demi\beta_{11} \right\},
}
where $\bS_+^2$ is the set of positive semidefinite matrices of order two. Define the convex cost function $F$ as follows:
\eq{\label{eq:cost_function_joint}
  F(t, x, \alpha,\beta) = \left\{ \begin{array}{ll} \displaystyle\sum_{i,j=1}^2 (\beta_{ij} - \bar\beta_{ij})^{2}   & \mbox{if } (\alpha,\beta)\in\Gamma, \\
		 +\infty & \mbox{otherwise,} \end{array} \right.
}
where $\bar\beta$ is a matrix of some reference values for $\beta$. Note that $\bar\beta$ may depend on $(t,x)$ as well.

The calibration instruments we consider are SPX European options, VIX options and VIX futures. The market prices of these products can be imposed as constraints on $X$. Let $G$ be a vector of $m$ number of SPX option (discounted) payoff functions. For example, if the $i$-th option is a put option with a strike $K_i>0$, then the payoff function $G_i:\R^2\to\R$ is given by $G_i(x) = \max(K_i-\exp(x_1), 0)$. Let $u^{SPX}\in\R^m$ be the market SPX option prices and $\tau\in[0,T]^m$ be the vector of their maturities. The prices $u^{SPX}$ can be imposed on $X$ by restricting $\bP$ to probability measures that satisfy
\eqn{
  \bE^\bP G_i(X_{\tau_i})=u^{SPX}_i,\qquad \forall i=1,\ldots,m.
}

Let $t\in[0,T]$. The annualised realised variance of the SPX price $S_t:=\exp(X^1_t)$ over a time grid $t_0<t_1<\cdots<t_n=T$ is defined to be
\eqn{
  AF\sum_{i=1}^{n}\left(\log\frac{S_{t_i}}{S_{t_{i-1}}}\right)^2,
}
where $AF$ is an annualisation factor. For example, if $t_i$ corresponds to the daily observation dates, then $AF=100^2\times252/n$, and the realised variance is expressed in basis points per annum. As $\sup_{i=1,\ldots,n}|t_i-t_{i-1}|\to 0$, the realised variance can be approximated by the quadratic variation of $X^1_t$, given by 
\eqn{
  AF\sum_{i=1}^n\left(\log\frac{S_{t_i}}{S_{t_{i-1}}}\right)^2 \overset{\bP}{\to} \frac{100^2}{T-t_0}\int_{t_0}^T\sigma_t^2\,dt.
}
The VIX index at $t_0$ is defined using a synthetic log-payoff option. In our setting, it can be equivalently re-written as the square root of the expected realised variance over the next 30 days (i.e., $T - t_0 = 30$ days), that is
\eqn{
  VIX_{t_0} &= \sqrt{\bE^\bP\bigg( \frac{100^2}{T-t_0}\int_{t_0}^T\sigma_t^2\,dt \,\bigg|\, \cF_{t_0} \bigg)} = 100\sqrt{\frac{2}{T-t_0}X^2_{t_0}}.
}

Consider VIX options and futures both with maturity $t_0$. Let $u^{VIX,f}\in\R$ be the market VIX futures price and let $u^{VIX}\in\R^n$ be the market VIX option prices. Let $H$ be a vector of $n$ number of VIX option payoff functions. Similarly to $G$, if the $i$-th VIX option is a put option with a strike $K_i>0$, then the payoff function $H_i:\R\to\R$ is given by $H_i(x)=\max(K_i-x,0)$. Let $J:\R^2\to\R$ be given by $J(x) := 100\sqrt{2x_2/(T-t_0)}$. Then, we want to further restrict $\bP$ to those under which $X$ also satisfies the following constraints:
\eqn{
  \bE^\bP J(X_{t_0}) &= u^{VIX,f}, \\
  \bE^\bP (H_i\circ J)(X_{t_0}) &= u_i^{VIX},\qquad \forall i=1,\ldots,n.
}

Finally, to ensure that $X^2_{T,T}=0$, one additional constraint is imposed on the model. Let $\xi:\R^2\to\R$ be a function such that $\xi(x) = 0$ if and only if $x_2=0$. Here, we choose $\xi(x):= 1-\exp(-(x_2)^2)$ and add constraint $\bE^\bP\xi(X_T)=0$. This constraint can be interpreted as a contract that has a payoff $\xi(X_T)$ at time $T$, and its price is always null. We will call it the \emph{singular contract}. 

We assume that $X_0=(X^1_0, X^2_{0,T})\in\R^2$ is known, and the initial marginal of $X$ is a Dirac measure on $X_0$. The value of $X^1_0$ is the logarithm of the current SPX price. In practice, $X^2_{0,T}$ can be inferred if the market prices of SPX call and put options maturing at $T$ are available over a continuous spectrum of strikes:
\eqn{
  X^2_{0,T} = \bE^\bP\left(\demi\int_0^T\sigma^2_s\,ds\right) = \int_0^{\hat f}\frac{\bE^\bP(k-S_T)^+}{k^2}\,dk + \int_{\hat f}^\infty\frac{\bE^\bP(S_T-k)^+}{k^2}\,dk,
}
where $\hat{f}=\bE^\bP (S_T)$ is the $T$-forward price of the SPX index (e.g., see \cite{carr1998volatility}). If $X^2_{0,T}$ is not observable from the market, we can treat it as a parameter. 

Now, to group all constraints together, we define 
\eqn{
  \begin{array}{r @{} c @{} c @{} c @{} c @{} c @{} c @{} c @{} c}
    c := (&u^{SPX}_1,\ldots,u^{SPX}_m &,&\; u^{VIX}_1,\ldots,u^{VIX}_n &,&\; u^{VIX,f} &,&\; 0 &), \\
    \cT := (&\tau_1,\ldots,\tau_m &,&\; t_0, \ldots,t_0 &,&\; t_0 &,&\; T &), \\
    \cG := (&\underbrace{G_1,\ldots,G_m}_{\text{$m$ SPX options}} &,&\; \underbrace{H_1\circ J, \ldots, H_n\circ J}_{\text{$n$ VIX options}} &,&\; \underbrace{J}_{\text{VIX futures}} &,&\; \underbrace{\xi}_{\text{singular contract}} &).
  \end{array}
}
Then the joint calibration problem can be reformulated as solving
\eq{\label{eq:main_joint_calibration}
  V_{joint} := \inf_{\bP\in\cP(X_0, c, \cT, \cG)}\bE^\bP\int_0^T F(t,X_t,\alpha^\bP_t,\beta^\bP_t)\,dt,
}
where
\eqn{
  \cP(X_0, c, \cT, \cG):=\{ \bP\in\cP^1 : \bP\circ X_0^{-1}=\delta_{X_0} \mbox{ and } \bE^\bP \cG_i(X_{\cT_i})=c_i ,\, i=1,\ldots,m+n+2 \}.
} 
Applying the arguments developed in Section 2, we can derive a dual formulation of \eqref{eq:main_joint_calibration}:
\eqn{
  V_{joint} = \sup_{\lambda\in\R^{m+n+2}}\lambda\cdot c - \phi(0,X_0),
}
where $\phi$ is a solution to the following HJB equation (in the viscosity sense):
\eqn{
  \dt\phi + \sup_{\beta\in\bS_+^2}\bigg\{ &-\demi\beta_{11}\partial_{x_1}\phi -\demi\beta_{11}\partial_{x_2}\phi + \demi\beta_{11}\partial_{x_1x_1}\phi + \demi\beta_{22}\partial_{x_2x_2}\phi \\
  & + \beta_{12}\partial_{x_1x_2}\phi - \sum_{i,j=1}^2(\beta_{ij}-\bar\beta_{ij})^2 \bigg\} = -\sum_{i=1}^{m+n+2}\lambda_i G_i\delta(t-\cT_i),
}
with the terminal condition $\phi(T,\cdot) = 0$.

\subsubsection*{Numerical example}

In \cite{guo2020joint}, the process $X$, also called the \emph{OT-calibrated} model, was calibrated to market data as of September 1st, 2020. The data consists of monthly SPX options maturing at 17 days and 45 days and monthly VIX futures and options maturing at 15 days. We also add the singular contract as a calibrating instrument (i.e., $\bE^\bP\xi(X_T)=0$) to ensure that the additional property $X^2_{T,T}=0$, $\bP$-a.s. is satisfied. 

Define $A(t,\kappa):=(1-e^{-\kappa(T-t)})/\kappa$ and $\nu(t,x,\kappa,\theta):= A(t,\kappa)^{-1}(2x - \theta(T-t)) + \theta$. The $\bar\beta$ in \eqref{eq:cost_function_joint} was set to
\eq{\label{eq:reference_heston_joint}
\bar\beta(t,x) = \left[ \begin{array}{cc} \nu(t,x_2,\bar\kappa,\bar\theta) & \demi\bar\eta\bar\omega A(t,\bar\kappa)\nu(t,x_2,\bar\kappa,\bar\theta) \\ \demi\bar\eta\bar\omega A(t,\bar\kappa)\nu(t,x_2,\bar\kappa,\bar\theta) & \frac{1}{4}\bar\omega^2 A(t,\bar\kappa)^2 \nu(t,x_2,\bar\kappa,\bar\theta) \end{array} \right],
} 
where $(\bar\kappa, \bar\theta, \bar\omega, \bar\eta) = (4.99, 0.038, 0.52, -0.99)$. The $\bar\beta$ in \eqref{eq:reference_heston_joint} was derived by reformulating a standard Heston model in terms of $X^1$ and $X^2$ defined in \eqref{eq:dynamics_joint_x1} and \eqref{eq:dynamics_joint_x2}. The parameters $(\bar\kappa, \bar\theta, \bar\omega, \bar\eta)$ have the usual interpretations as in the Heston model and are obtained by (roughly) calibrating a Heston model to the SPX option prices. The initial position is $X_0 = (8.1673, 0.0048)$. In addition, a \emph{reference iteration} method was used for smoothing the volatility surfaces and skews, which is similar to iterating $\bar\sigma$ in the local volatility example presented in Section \ref{sec:local_vol}. We refer the reader to \cite{guo2020joint} for more details.

\begin{figure}[ht!]
   \begin{minipage}{0.49\textwidth}
     \centering
     \includegraphics[width=1.0\linewidth]{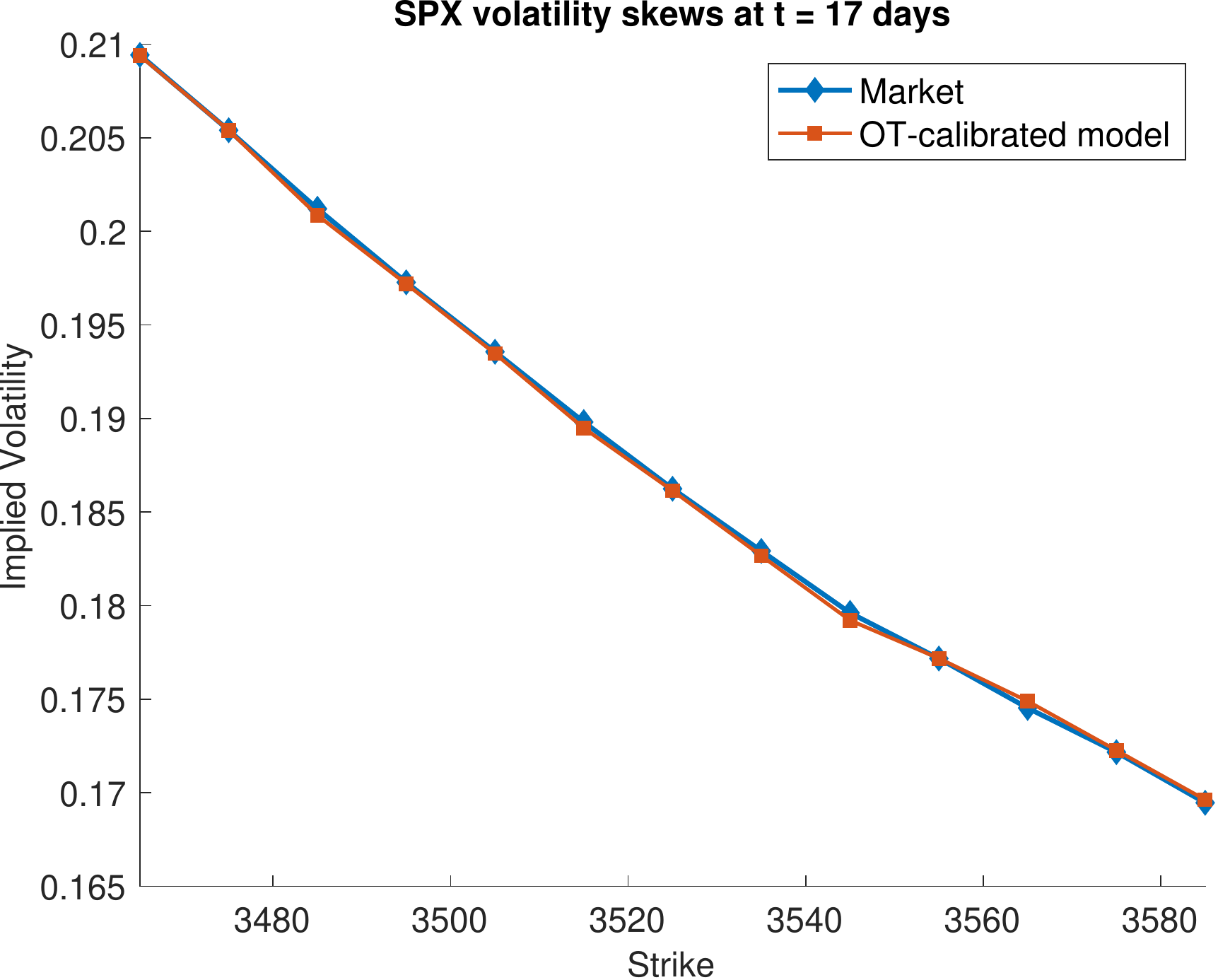}
   \end{minipage}\hfill
   \begin{minipage}{0.49\textwidth}
     \centering
     \includegraphics[width=1.0\linewidth]{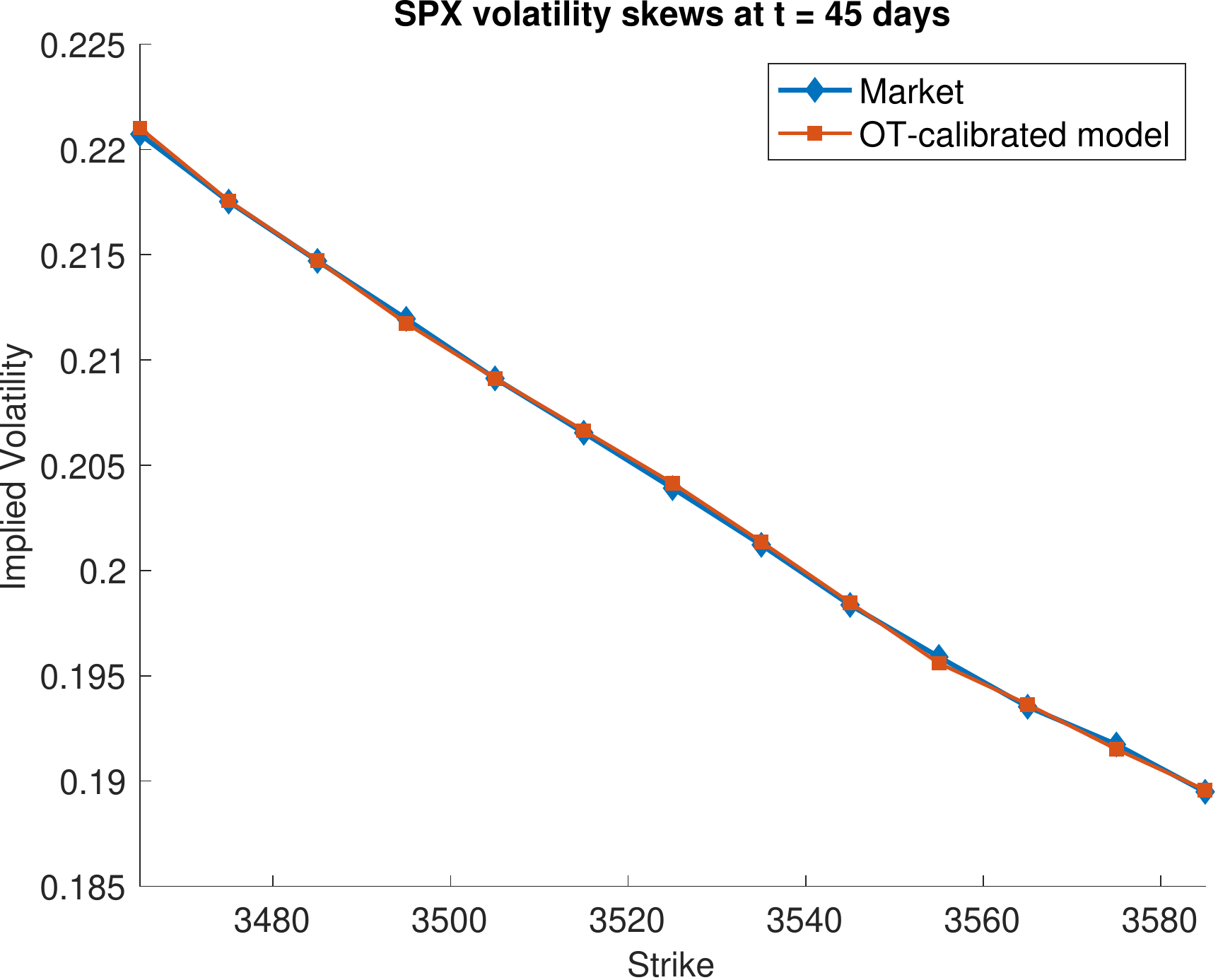}
   \end{minipage}
   \centering
   \begin{minipage}{0.49\textwidth}
     \centering
     \includegraphics[width=1.0\linewidth]{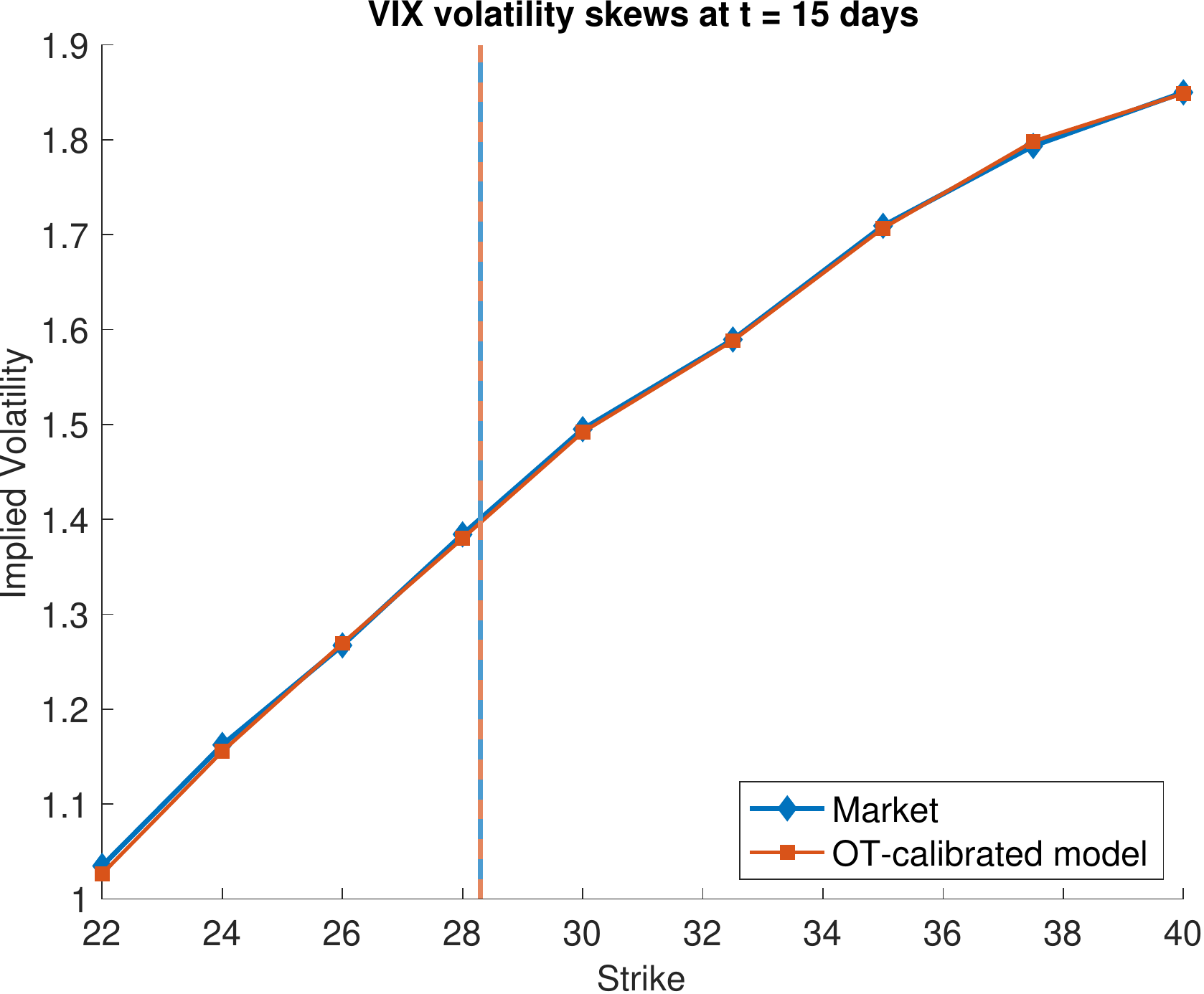}
   \end{minipage}
   \caption{Approximated OT-calibrated model volatility skews of SPX options at $17$ days, SPX options at $45$ days and VIX options at $15$ days in the joint calibration numerical example. The vertical lines are VIX futures prices. Markers correspond to computed prices which are then interpolated with a piece-wise linear function.}
   \label{fig:IV_result_joint}
\end{figure}

\begin{figure}[ht!]
\begin{center}
\includegraphics[width=1.0\textwidth]{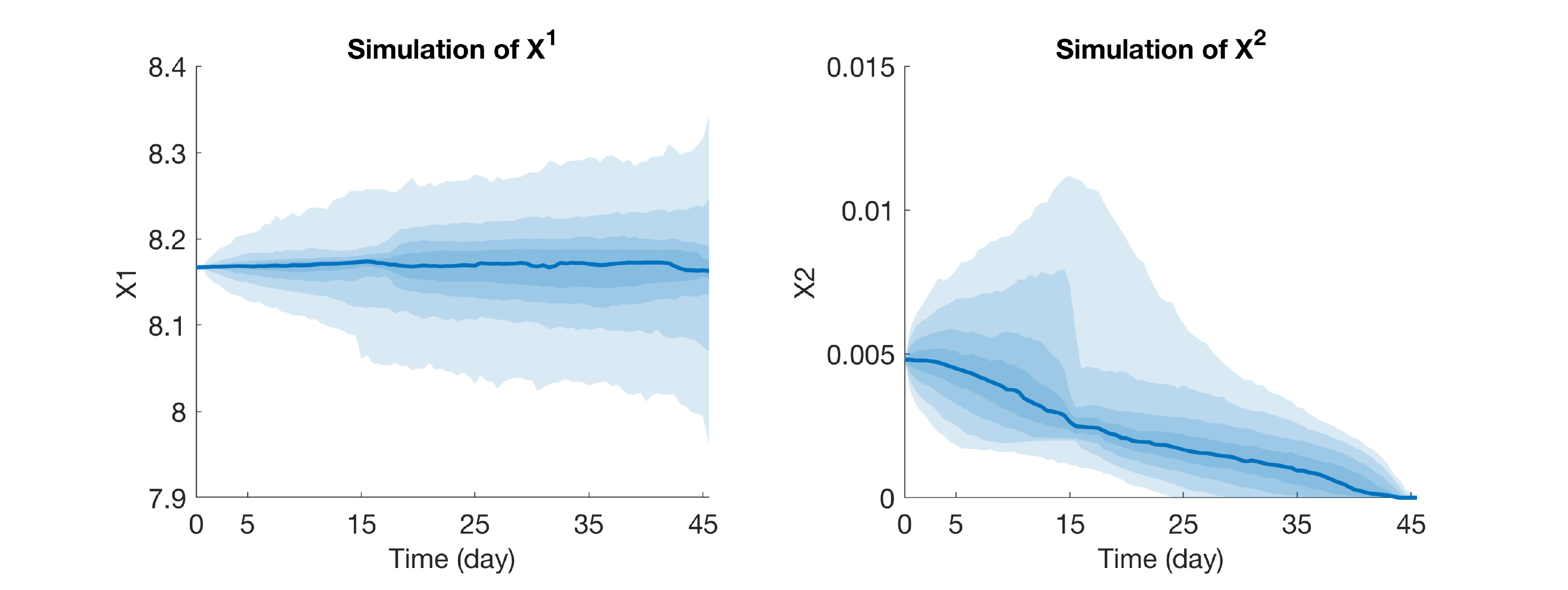}
\caption{The simulations of the OT-calibrated model $X$ in the joint calibration numerical example. }
\label{fig:simulation_X_joint}
\end{center}
\end{figure}

Figure \ref{fig:IV_result_joint} shows the model volatility skews of the OT-calibrated model. The simulation of $X$ is given in Figure \ref{fig:simulation_X_joint}. The results show that the model accurately attains the market prices while keeping the property $X^2_{T,T}=0$, $\bP$-a.s. satisfied. 
 
\subsection{Path-dependent volatility calibration}\label{sec:path-dependent}

The result of \cite{guo2018path} can be applied to the calibration of volatility models to path-dependent options. In \cite{guo2018path}, a path-dependent model, which has the same form of \eqref{eq:dynamics_lv}, was first calibrated to European options and then to path-dependent options. The theoretical developments are out of the scope of this paper, so we only highlight the results here.

Consider probability measures $\bP$, with sufficient regularity, under which $X$ takes the form of \eqref{eq:dynamics_lv}. We also consider the same cost function defined in Section \ref{eq:costfun_lv}, which is \eqref{eq:costfun_lv}. If we only consider European options as the calibrating instruments, the path-dependent model recovers a local volatility model, and the calibration problem is equivalent to the one introduced in Section \ref{sec:local_vol}. When calibrating to path-dependent options, instead of solving the HJB equations \eqref{eq:hjb_lv}, one needs to solve a class of path-dependent PDEs (PPDE) which is numerically difficult to solve. By identifying the relevant path-dependent state variables, the infinite dimensional PPDE reduces to a finite dimensional PDE which depends on the spot price as well as the additional path-dependent state variable and can be solved via conventional numerical methods. Here are some examples of relevant state variables:
\begin{itemize}
  \item European options: the spot price $X_t$;
  \item Asian options: the spot price $X_t$ and the running average $\frac{1}{t}\int_0^t X_t\,dt$;
  \item Continuous barrier options: the spot price $X_t$ and the indicator variable $\mathds{1}(X_s>B,s\in[0,t])$ for lower barriers or $\mathds{1}(X_s<B, s\in[0,t])$ for upper barriers;
  \item Lookback options: the spot price $X_t$ and either the running minimum $\min_{s\in[0,t]}X_s$ or running maximum $\max_{s\in[0,t]}X_s$.
\end{itemize}

\subsubsection*{Barrier options}

Here, we give an example of barrier options (see \cite{guo2018path} for more examples). Formally speaking, a barrier is a closed subset $\bB\subset[0,1]\times\R$ whose complement is a connected region containing $(0,X_0)$. The payoff of a barrier product expiring at time $T$ is a function of $X_T$ and the indicator variable $\mathbf{1}_t:=\mathds{1}(X_s\in \bB, \mbox{for some } s\in[0,t])$, checking whether the path of the underlying has hit the barrier. When calibrating to a collection of barrier products with a single fixed barrier, the required state variables are $t, X_t$ and $\mathbf{1}_t$. Then the function $\phi$ can be effectively split into two functions, $\phi^0(t,x)$ and $\phi^1(t,x)$, corresponding to the cases $\mathbf{1}_t=0$ and $\mathbf{1}_t=1$, respectively. The dual formulation of the calibration problem is solving
\eqn{
  \sup_{\lambda\in\R^m} \lambda\cdot c - \phi(0,X_0),
}
where $\phi$ is the solution to the split PDE
\eqn{
  \dt\phi_1 + F^*\left(\dx\phi_1,\demi\dxx\phi_1\right) = 0, &\quad (t,x) \in [0,1]\times\R, \\
  \dt\phi_0 + F^*\left(\dx\phi_0,\demi\dxx\phi_0\right) = 0, &\quad (t,x) \notin \bB, \\
   \phi_0 = \phi_1, &\quad (t,x) \in \partial\bB,
}
where $F^*$ is the convex conjugate of $F$ defined in \eqref{eq:costfun_lv} with respect to the last two variables. Similarly, the optimal volatility will be switching between two local volatilities $\sigma_0(t,x)$ and $\sigma_1(t,x)$, conditional to whether the underlying has hit $\bB$ or not. The PDE for $\phi_0$ will be used to compute the volatility function prior to the stock hitting the barrier, while the PDE for $\phi_1$ will be used to compute the volatility function after the barrier has been hit.

\subsubsection*{Numerical example}

\begin{figure}[ht!]
  \begin{minipage}{0.49\textwidth}
    \centering
    \includegraphics[width=1.0\linewidth]{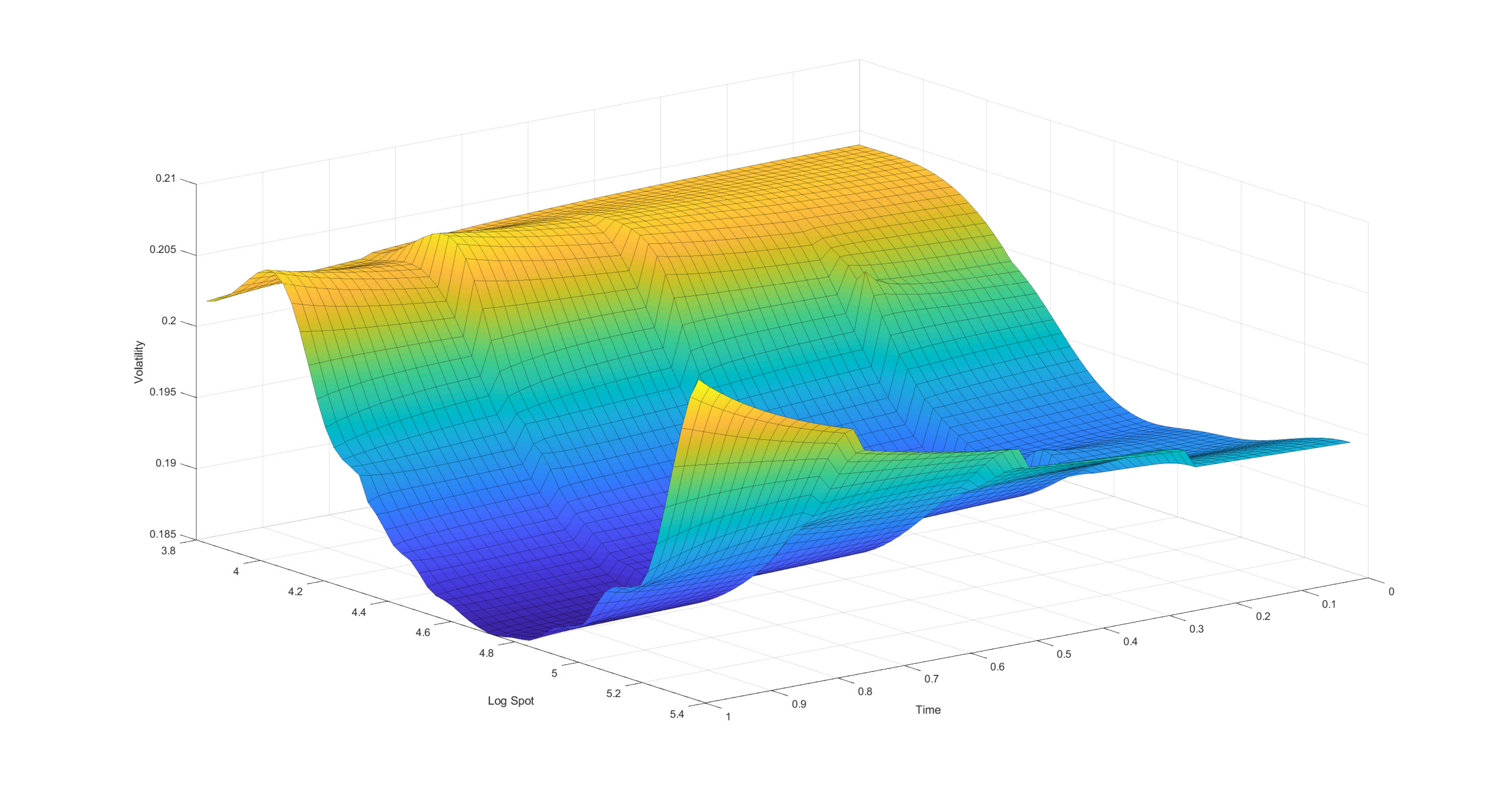}
  \end{minipage}\hfill
  \begin{minipage}{0.49\textwidth}
    \centering
    \includegraphics[width=1.0\linewidth]{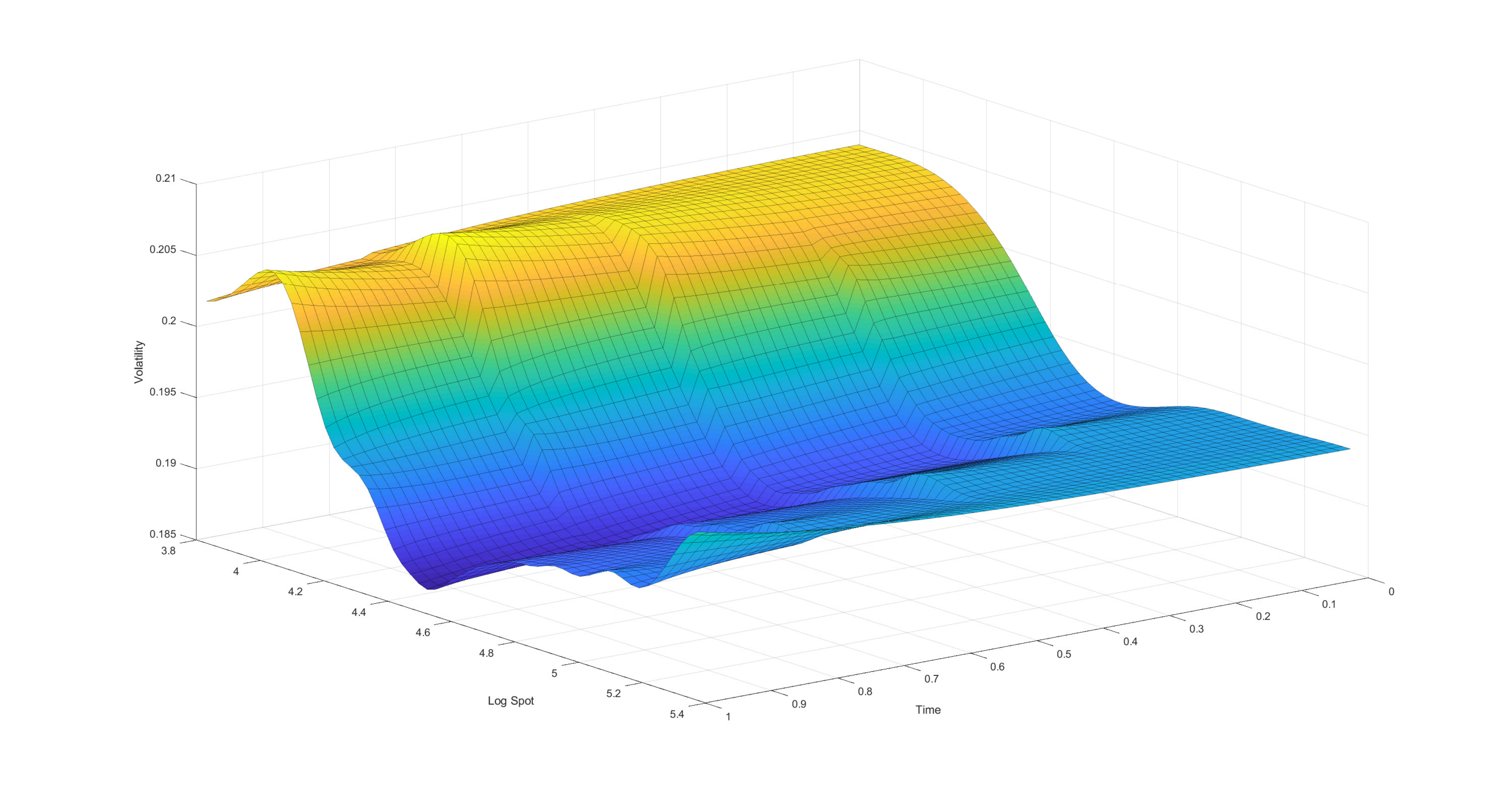}
  \end{minipage}
  \caption{Volatility calibrated to all down-and-in and down-and-out puts with strikes at all the grid points and four different maturities. The left half is showing $\sigma_0$ (before hitting the barrier) and the right half is showing $\sigma_1$ (after hitting the barrier).}
  \label{fig:path_vol_barrier}
\end{figure}

\begin{figure}[ht!]
\begin{center}
\includegraphics[width=1.0\textwidth]{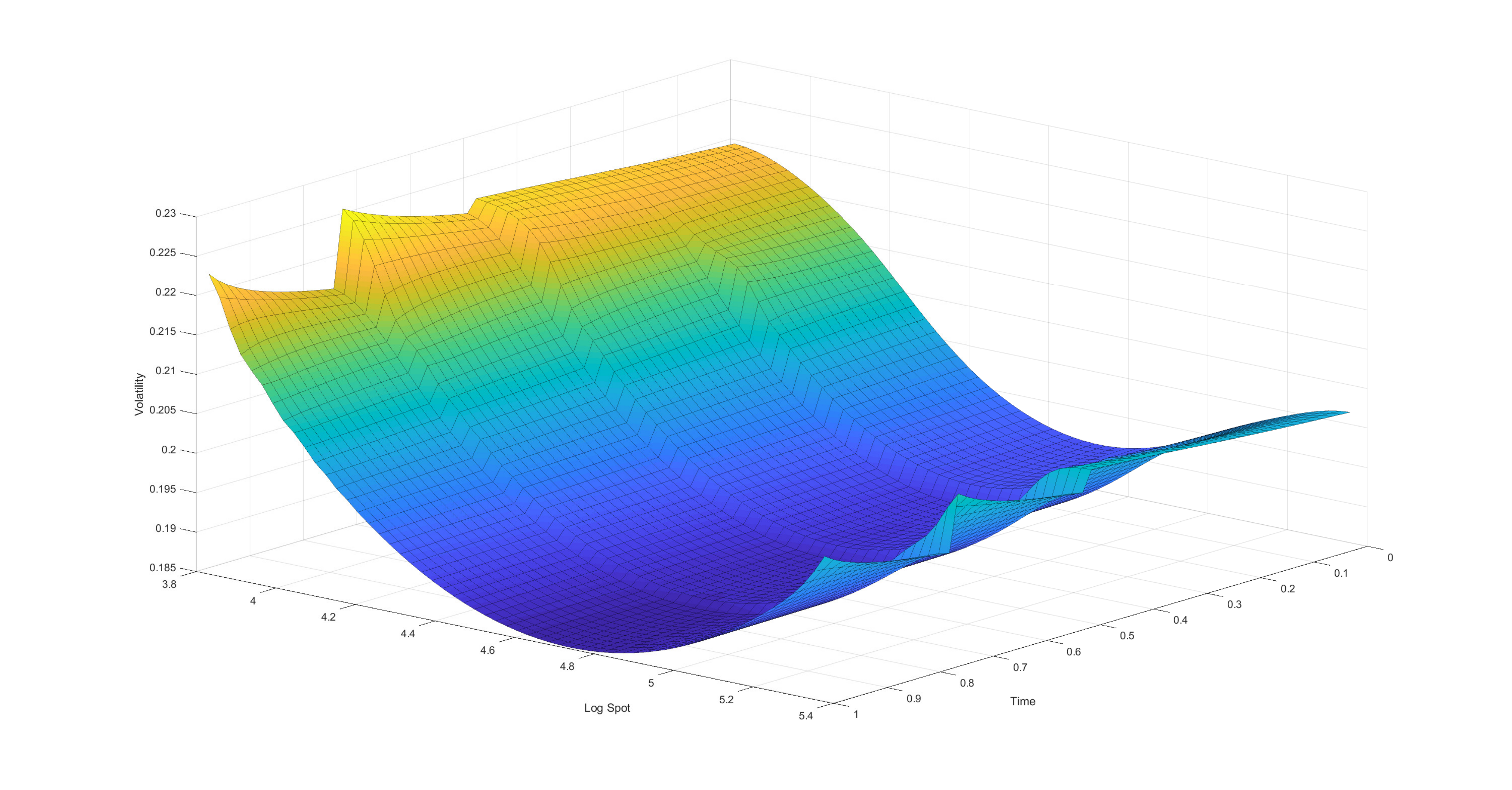}
\caption{Volatility surface calibrated to European put options with strikes at all the grid points and four different maturities.}
\label{fig:path_vol_euro}
\end{center}
\end{figure}

As an example, let us consider barrier products with respect to a continuous lower barrier $\{x\leq b\}$ where $b<X_0$ is a constant. In particular, we will be calibrating to all down-and-in and down-and-out puts with strikes at all the grid points and four different maturities. The left half of Figure \ref{fig:path_vol_barrier} shows the calibrated volatility function $\sigma_0$ (before hitting the barrier) and the right half shows $\sigma_1$ (after hitting the barrier). Even though $\sigma_0$ is only defined for $x\geq b$, for the purpose of visualisation, we set $\sigma_0=\sigma_1$ for $x<b$. For comparison, the volatility calibrated to only European options with the same strikes and maturities is shown in Figure \ref{fig:path_vol_euro}.

\bibliography{biblio-greg,biblio-leo}
 
\end{document}